\documentclass[14pt]{article}
\pdfoutput=1
\usepackage{amsmath,amssymb,amsfonts,amsthm,bm,bbm,cancel,wasysym}
\usepackage{epsfig,graphics,graphicx,epstopdf}
\usepackage{array,booktabs,colortbl,colordvi,multirow}
\usepackage{colordvi,color,xcolor}
\usepackage{hyperref}
\usepackage{rotating}

\usepackage{verbatim}
\usepackage{cite}
\usepackage{subfig}
\usepackage{setspace}
\usepackage{url}
\usepackage[percent]{overpic}
\usepackage{slashed}
\usepackage{authblk}
\usepackage{xspace}
\usepackage{fullpage}
\usepackage{hyperref}

\def\ie{{\it i.e.}}
\def\eg{{\it e.g.}}

\def\etal{{\it et al.}}

\def\to{\rightarrow}

\newskip\zatskip \zatskip=0pt plus0pt minus0pt
\def\matth{\mathsurround=0pt}
\def\lsim{\mathrel{\mathpalette\atversim<}}
\def\gsim{\mathrel{\mathpalette\atversim>}}
\def\atversim#1#2{\lower0.7ex\vbox{\baselineskip\zatskip\lineskip\zatskip
  \lineskiplimit 0pt\ialign{$\matth#1\hfil##\hfil$\crcr#2\crcr\sim\crcr}}}

\begin{document}


\begin{flushright}
SLAC-PUB-17326\\
\today\\
\end{flushright}
\vspace*{5mm}

\renewcommand{\thefootnote}{\fnsymbol{footnote}}
\setcounter{footnote}{1}

\begin{center}

{\Large {\bf Kinetic Mixing and Portal Matter Phenomenology}}\\

\vspace*{0.75cm}

{\bf Thomas G. Rizzo}~\footnote{rizzo@slac.stanford.edu}

\vspace{0.5cm}

{SLAC National Accelerator Laboratory}\ 
{2575 Sand Hill Rd., Menlo Park, CA, 94025 USA}

\end{center}
\vspace{.5cm}

\begin{abstract}
 
\noindent

Dark photons (DP) are interesting as potential mediators between the dark matter (DM) sector and the fields of the Standard Model (SM). The interaction of the dark photon, described 
by a broken $U(1)_D$ gauge symmetry, with the SM is usually generated at the one-loop level via kinetic mixing through the existence of portal matter (PM), here assumed to 
be fermionic,  which carries both a dark charge as well as a SM $U(1)_Y$ hypercharge. For theoretical consistency, as well as for many phenomenological reasons, this portal matter must 
be vector-like with respect to the SM and dark gauge groups and, in particular, is shown to be allowed only to transform as vector-like copies of the usual SM fields. The dark Higgs 
that is responsible for the breaking of $U(1)_D$ can then generate a mixing between the portal matter and SM fields with the same electric charge thus altering the dark photon/portal 
matter interactions with (at least some of) the SM fields and also providing a path for the portal matter fields to decay. In this paper we briefly explore the phenomenology of 
some specific simple models of this portal matter including, for the case where the portal matter is leptonic in nature, their potential impact on experiments probing low energy 
parity-violation and the g-2 of the muon. In the case of color-triplet, bottom quark-like portal matter, their direct pair- and single-production at the LHC is shown to be observable in final 
states that include missing $E_T$ and/or very highly boosted lepton-jets together with pairs of high-$p_T$ b-jets that can be used to trigger on such events. These signatures are quite 
distinct from those usually employed in the search for vector-like quarks at the LHC and, furthermore, we demonstrate that the conventional signal channels for vector-like quarks 
involving the SM Higgs and gauge fields are essentially closed in the case of portal matter. Many other more complex, and more realistic, portal matter scenarios of the type discussed 
here are possible which can lead to wide-ranging signatures in various classes of experiments.
\end{abstract}

\renewcommand{\thefootnote}{\arabic{footnote}}
\setcounter{footnote}{0}
\thispagestyle{empty}
\vfill
\newpage
\setcounter{page}{1}



\section{Introduction}

The nature of dark matter (DM) remains as one of the great mysteries facing particle physics. For DM to achieve its observed abundance, it is probable that it must have some new, 
non-gravitational interaction(s) with the fields of the Standard Model (SM) - but such interactions are likely to be far weaker than the usual ones with which we are familiar. Traditionally, 
both axions\cite{Kawasaki:2013ae,Graham:2015ouw} and WIMPs\cite{Arcadi:2017kky} have been the leading candidates for DM, originating in top-down frameworks that addressed 
other issues such as the strong CP problem and naturalness. Searches for such particles have so far obtained only null results and while these important experiments are still evolving the 
set of possible DM candidates continues to grow as does the possible ways to look for them\cite{Alexander:2016aln,Battaglieri:2017aum}. Among these, one of the most interesting 
and directly testable possibilities that has garnered significant recent attention is the kinetic mixing/vector portal model\cite{vectorportal,KM}. In its most simple form, this setup assumes 
the existence of a dark photon (DP), \ie, the gauge field of a new $U(1)_D$, that is spontaneously broken through the vacuum expectation value (vev) of a complex SM singlet dark 
Higgs field, $<S>=v_s/\sqrt{2}$, which here we will assume to lie in the range $v_s \sim 0.1$ to a few GeV.  The DM field then carries a non-zero dark charge, $Q_D$, so that it can 
couple directly to the dark photon. On the other hand, the SM fields do not carry any dark charges under this new $U(1)_D$ so that the interactions of the dark photon with the SM fields 
are loop-induced via the kinetic mixing (KM) of the dark photon with the usual 
SM $U(1)_Y$ hypercharge field, $B_{\mu}$, whose mixing strength is characterized by a parameter, $\epsilon \sim 10^{-(3-5)}$ for dark photons of mass $\sim 0.1-1$ GeV or so. The 
vacuum polarization-like diagram that generates this interaction is the result of the existence 
of an important (though often neglected) new set of fields which carry {\it both} SM as well as dark charges; we will refer to such dual-charged states as portal matter in 
what follows. This portal matter thus plays an essential role in all models where the DM interactions with the SM are induced by KM. As will be further discussed below, assuming that the KM 
is indeed generated only at one-loop to have a phenomenologically interesting strength, in order to obtain a non-zero and yet finite result (as might be expected in 
UV-complete scenarios) several, non-degenerate portal matter fields are required, the product of whose charges must satisfy the condition $\sum_i Q_{Y_i}Q_{D_i}=0$. What are these 
portal matter states and how do they impact non-DM phenomenology? These portal  
matter fields which form a necessary ingredient of this setup have so far received very little attention much less detailed study in the literature\cite{however} particularly in the case where they 
carry non-trivial SM $SU(3)_c$ charges. To rectify this situation, it is the 
nature of these portal matter fields, their interactions with the dark photon, as well as with the familiar SM particles that will be of interest to us below. We will see 
that even after the obligatory constraints 
are satisfied there remains considerable flexibility in the identity and spectrum of the portal matter states which can result in broad phenomenological implications which we can only 
begin to examine here. How these portal matter states may fit into a more complete theoretical structure will be discussed elsewhere.

If the portal matter fields are fermionic, which we will argue below is the most likely scenario and which we will assume in the following discussion, they must be vector-like with respect to both 
the $U(1)_D$ and the SM gauge groups in order to avoid various gauge anomalies,  precision electroweak constraints and the additional constraints arising from the Higgs boson's  
production and decay properties as measured at the LHC. Such vector-like 
fermions (VLF) are then qualitatively similar in nature to those that have been well-studied in various extensions of the SM\cite{vlf} and which have been sought at the LHC without success 
(so far) by both the ATLAS and CMS collaborations\cite{searches}.  The difference in the present case is that these portal matter fields also carry dark charges so that they cannot mix with the 
correspondingly charged SM fields via a tree-level Yukawa coupling with the SM Higgs. This implies that their decays to final states involving either the SM Higgs or the SM gauge bosons 
are highly suppressed yet these are the very modes being searched for at the LHC. However, such mixings between the portal matter and the fermion fields of the SM {\it can} occur 
when the $U(1)_D$ charges 
of these portal matter fields allow for an analogous Yukawa coupling to the {\it dark} Higgs which, as we'll see below, leads to quite a different phenomenology and production signatures at 
the LHC now involving long-lived states arising from the dark photon and dark Higgs. Furthermore, we will show that such mixings are required in order to satisfy numerous 
lifetime constraints on 
such new particles carrying SM charges. As we will discuss, the existence of such states can also have direct impact on other experiments performed 
at lower energies depending upon their specific SM quantum numbers. Much of this phenomenology, as we will see, has been previously encountered in dark photon models in other contexts 
at the qualitative level. However, its specific nature and origin in the present analysis, as we will see below, are quite different quantitatively from that most frequently encountered 
in the literature. 

The outline of this paper is as follows:  Section 2 begins by providing a general overview of the KM/dark photon setup and the role that portal matter plays within it. After specializing to the case of the 
portal matter being a set of SM-like vector-like fermion, the mixing of these portal matter fields with those of the SM are analyzed in detail in the most minimal scenario involving only a single SM flavor 
pair of portal matter fields. As noted above, much of this phenomenology will be familiar at the qualitative level but will shown to generally differ quantitatively from these other more 
well-known results. Section 3 provides 
a non-comprehensive survey of some of the possible phenomenological implication of the existence of the mixing between the SM fields and the portal matter but which depend on the detailed 
SM quantum numbers assumed for the portal matter states. The modifications of the dark photon couplings to the SM fields that are induced by this mixing are then considered; it is pointed out that 
these, generally parity-violating, contributions can be comparable in strength to those induced by the familiar KM. The portal matter/SM mixing also induces off-diagonal couplings of these 
portal matter/SM fields to both the dark photon and the dark Higgs. In the case of color-singlet portal matter, if 
the portal matter transforms like a SM right-handed electron then such new parity-violating effects may be observable in low energy 
experiments such as MOLLER. If the portal matter is $\mu_R$-like in nature, this leads to multiple new contributions to the g-2 of the muon, whose experimental value is known to differ from 
the SM prediction by $\sim 4\sigma$.  We note that if the portal matter were similar to the left-handed lepton doublet, then the mixing with the vector-like fermion $N$ induces a new low-$Q^2$ interaction 
between neutrinos and quarks which may impact measurements at, \eg, MiniBooNE or DUNE. Lastly, and in more detail, we consider the setup where the portal matter transforms as a color triplet, 
in particular like $b_R$ in the SM; 
this is the main emphasis of the current work. In such a case a variety of possible signals can arise from their pair production at the LHC. In addition to the two, high-$p_T$, $b$-jets 
produced in portal matter pair decay which can be employed as a tag, the final state is shown 
to include missing $E_T$, producing a SUSY-like signature,  and/or displaced lepton-jets which are atypical due their very large boosts. The case where the $b_R$-like portal matter is singly 
produced is also briefly discussed,  is shown to lead to asymmetrically boosted lepton-jets and to potentially have a significant production rate at the LHC.  A comparison to the case 
where the portal matter transforms as $t_R$-like is also briefly made. Section 4 contains a final discussion and our conclusions. An Appendix summarizes the most important couplings 
appearing in the present analysis.


\section{Portal Matter Phenomenology Setup}

In this paper we will be particularly interested in the range of dark matter and dark photon masses $\sim 0.1-1$ GeV and below that will be accessible in, \eg, future fixed target and
light DM direct search experiments.  
Within this context, consider a set of vector-like fermions, $F_i$, with masses $m_i$ and whose hypercharges and corresponding dark charges are denoted by $Q_{Y_i}$ and 
$Q_{D_i}$, respectively. With KM absent 
at tree-level, then, in conventional normalization\cite{KM}, the familiar 1-loop vacuum polarization-like graph coupling the two corresponding gauge fields will lead to experimentally 
interesting values\cite{Alexander:2016aln,Battaglieri:2017aum} for the KM parameter $\epsilon \sim 10^{-(3-5)}$ given by ($c_W=\cos \theta_W$)\cite{first} 

\begin{equation}
\epsilon =c_W \frac{g_Dg_Y}{12\pi^2} \sum_i ~Q_{Y_i}Q_{D_i}~ ln \frac{m^2_i}{\mu^2}\,
\end{equation}
where obtaining a finite result (as might be expected in UV-complete scenarios) requires that these fields be non-degenerate and satisfy the constraint 
$\sum_i Q_{Y_i}Q_{D_i}=0$; here $g_{Y,D}$ are the hypercharge and dark gauge coupling constants, respectively. We will assume that 
$g_D \sim 0.1-0.3$ in the numerical analysis that follows. The very simplest possibility, which we will consider here, is thus where there are only two such vector-like fermion, $F_{1,2}$, with different 
masses,  having the same value of the SM hypercharge but with simultaneously opposite values for $Q_D${\footnote {The opposite charge assignment possibility provides some 
interesting model building challenges if the portal matter transforms in a non-trivial manner under the SM $SU(2)_W$ or $SU(3)_c$ gauge groups.}}; we will take this numerical value to be 
unity, $|Q_{D_i}|=1$, without loss 
of generality in the present analysis. Thus, in this limited discussion, we will constrain ourselves to the case where the portal matter consists of a single pair of fields that have identical SM 
characteristics. Clearly, we can generalize far beyond this simplifying assumption to more complex, but perhaps more realistic and interesting, 
situations but we will leave this to later work.)  Given these considerations, we then obtain numerically
\begin{equation}
\epsilon \simeq 1.0\Big(\frac{g_D}{0.1}\Big) N_cQ_Y\frac{ln (m_2/m_1)}{ln ~1.5} \cdot 10^{-4}\,
\end{equation}
a value which is consistent with our expectations\cite{Alexander:2016aln,Battaglieri:2017aum} and where $N_c$ is the number of colors of the portal matter fields. Since these portal matter states 
carry both SM and dark charges they must also be {\it unstable} to conform with cosmological constraints and so a decay path for them must be generated in any realistic theory. In 
practice this implies that they must mix with the analogous SM fields as we will discuss further below. 

It should again be stressed that this rather simple scenario need not be the one that is actually realized in nature and many interesting combinations of the vector-like fermion fields playing 
the role of portal matter are possible. For example, one can imagine that in more UV-complete scenarios, \eg, portal matter contributions from color triplet, weak iso-singlets naturally 
`canceling' against those from color singlet, weak iso-doublets as in a $5+\bar 5$ SU(5)-like representation as happens in the case of KM in GUTS. Such possibilities may open a 
window to a more realistic UV completion 
of the present toy model than we contemplate here~\cite{tomorrow}. Indeed, such possibilities will be considered elsewhere but one should be 
mindful of their potential phenomenological impact which may lie simultaneously in multiple phenomenological sectors and having broad experimental implications. 
We note in passing that if we had instead chosen the portal matter to be complex scalars 
{\it without} vevs, it is much more difficult to arrange for them to decay via mixing with the analogous SM field, \ie, the Higgs, which is the only available SM scalar, without further 
augmentation of the model.

Once a value of $\epsilon \neq 0$ is obtained this leads to the now standard development\cite{vectorportal,KM} (which we now briefly review to set the stage for later discussion) via 
the Lagrangian
\begin{equation}
{\cal L}={\cal L}_1+{\cal L}_2+{\cal L}_{HS}+{\cal L}_{Portal}\,
\end{equation}
with the ${\cal L}_1$ piece of the action describing the dark photon field, $\hat V$, the corresponding hypercharge gauge field, $\hat B$, and the rest of the SM:
\begin{equation}
{\cal L}_1=  ~-\frac{1}{4} \hat V_{\mu\nu} \hat V^{\mu\nu} -\frac{1}{4} \hat B_{\mu\nu} \hat B^{\mu\nu} +\frac{\epsilon}{2c_w} \hat V_{\mu\nu} \hat B^{\mu\nu}  + {\cal L}_{SM}   \,,  
\end{equation}
and, with $D_\mu=\partial_\mu +ig_D Q_D \hat V_\mu$ being the gauge covariant derivative in obvious notation,
\begin{equation}
{\cal L}_{2}=   ~ i\bar \chi\gamma^\mu D_\mu \chi-m_D\bar \chi \chi + (D_\mu S)^\dagger (D^\mu S) +\mu_S^2 S^\dagger S -\lambda_S (S^\dagger S)^2 \,,
\end{equation}
which describes the DM field $\chi${\footnote {Additional terms may also be present in this piece of the Lagrangian to generate an additional Majorana mass for $\chi$ if desired.}}.  
Note that here the DM is assumed to be a fermion (which does not couple directly to the SM) for definiteness but the detailed nature of the DM will not 
play any significant role in the discussion that follows below and is included here only for completeness but will be more important in our subsequent work\cite{tomorrow}. 
This same piece of the action also describes the dark Higgs, $S${\footnote {Here we will often write $S$ as 
$(v_s+S+iG)/\sqrt 2$, where $G$ is the Goldstone boson eaten by the dark photon  
and $S$ now represents the remaining real scalar field.}}. We will make the further assumption in our discussion below that $2m_D>m_V$ such that $V$ must decay to SM fields, 
\eg, $e^+e^-$, so that this state may be observable at, \eg,  HPS\cite{Baltzell:2016eee} but not at, \eg,  LDMX\cite{LDMX} if one searches only for final states with missing 
energy/momentum. The KM in ${\cal L}_1$ is removed and the fields become canonically normalized via the standard field re-definitions (to lowest order in $\epsilon$), 
$\hat B_\nu \to B_\nu +\frac{\epsilon}{c_W}V_\nu$, $\hat V_\nu \to V_\nu$.  After this transformation and mass mixing with the SM $Z$ generated via the vev of the SM Higgs, $H$, is 
accounted for (in the limit of small $m^2_V/m^2_Z$  as is relevant in the present case), the physical $V$ mass eigenstate couples to the SM fields as $\simeq e\epsilon Q_{em}$. 
It should be noted that since $m_V=g_Dv_s$ with $g_D\sim 0.1-0.3$ as being assumed here while $S$ has a mass $\sqrt {2\lambda_S}v_s$ 
it is natural that $m_S>m_V$ since $g_D$ is small. Note that $S$ has no gauge-invariant, renormalizable coupling to the Dirac fermion DM. This implies, in the absence of any 
mixing with the SM Higgs, that $S$ decays as 
$S\to VV^*, V^*\to e^+e^-$ and possibly more likely via the on-shell process $S\to VV$ and not directly to DM states. In the $S\to VV$
case, this decay is prompt as we are assuming $g_D \sim 0.1-0.3$; this will be important in the discussion that follows.  

As well-known the $H$ and $S$ Higgs fields can also mix due to their vevs through the standard quartic term
\begin{equation}
 {\cal L}_{HS} = ~\lambda_{HS} H^\dagger H S^\dagger S \,
\end{equation}
with the magnitude of $\lambda_{HS}$ usually and necessarily being fine-tuned to very small values (unlike what can happen by employing extra 
dimensions\cite{Rizzo:2018ntg, Rizzo:2018joy} where this mixing can be naturally set to zero via a choice of boundary conditions) to meet phenomenological constraints arising from 
measurements of Higgs properties at the LHC.  To get an idea of this stringent requirement which will be important in later discussions, we can compare the mixing-induced 
partial width $\Gamma(H\to SS+VV)$ to the present limit on the invisible Higgs branching fraction $B=BF(H\to inv) \lsim 0.20-0.25$~\cite{inv}.  
To leading order in this mixing, the $HSS$ coupling is 
just $\simeq \lambda_{HS}v_H/2$ with $v_H \simeq 246$ GeV. Some algebra then tells us that this BF bound implies that 
$|\lambda_{HS}| \lsim 4.7\cdot 10^{-3}~ \sqrt{B/0.1}$. (Note that a bound of similar magnitude can be obtained by requiring that the mass of $S$ dominantly arises from its own vev.)
Since one finds that the $H-S$ mixing angle is then $\theta_{HS}\simeq -\lambda_{HS}v_s/2\lambda_Hv_H$,  this result 
further implies that $|\theta_{HS}|\lsim 1.8\cdot 10^{-4} ~(\frac{v_s/v_H}{10^{-2}})~ \sqrt{B/0.1}$. In the current analysis, except where noted, we will generally set the $\lambda_{HS}$ 
coupling to an 
extremely small value, corresponding to a very tight mixing angle bound, so that its influence on the dark sector fields can generally be safely ignored. However, we will return to this 
very important issue later on below.

So far we have briefly summarized the usual development of the dark photon/KM scenario to establish notation and basic assumptions; now let us return to a discussion of the portal 
matter part of the action. We recall that these vector-like portal matter fermion states must be unstable so that at least one path for their decay must exist. For the more `conventional' 
vector-like fermion that are usually discussed such decay paths are provided by their mixing (via the vev of the SM Higgs) with the analogous SM states which carry the same QCD 
and QED quantum numbers. For portal matter states this mixing option does not occur as they also 
carry dark charges while the SM Higgs does not and so a gauge-invariant coupling of the required type is absent. We can, however, arrange something quite similar instead by employing  
the SM singlet dark Higgs\cite{Davoudiasl:2012ig}. Amongst the many {\it a priori} possibilities to consider at the renormalizable level, since the dark Higgs is a SM singlet, the pair of 
portal matter fields, $F_{1,2}$ (with $Q_D=\pm 1$) {\it must} be chosen to transform as one of the familiar set of vector-like fermions (vector-like fermion), either vector-like quarks (VLQ) 
or vector-like leptons (VLL),  which are analogous to the representations already appearing in the SM, \ie, :
\begin{equation}
 ~~~ (T,B)^T, ~(N,E)^T, ~T, ~B, ~E \,
\end{equation}
where the simplest (but not necessary) possibility is clearly that the portal matter are transform just like one of the SM $SU(2)_W$ isosinglets, \ie, $T,B,E$. It is now easy to see that only 
these 5 particular choices allow for the new states to decay to SM fields at tree-level since $S$ is a SM singlet. To determine how this specific mixing may be generated we imagine that 
$|Q_D(S)|=1$;  then we can write a new gauge invariant term in the Lagrangian given by 
\begin{equation}
 {\cal L}_{Portal} = ~ \lambda_{1a} \bar F_{1L}f^a_RS +\lambda_{2a} \bar F_{2L}f^a_RS^\dagger +h.c. \,
\end{equation}
with $L(R)$ the usual helicity projections and $f^a$ being the corresponding SM fermion of the same color and hypercharge(\ie, electric charge) with family label $a=1,2,3$. When 
$S$ gets a vev, this piece of the action will generate off-diagonal mass terms, symbolically of the form $X^a_i=\lambda^a_i v_s/\sqrt {2}$, inducing a mixing between portal matter fields 
$F_{1,2}$  with those in the SM, $f^a$, and eventually between themselves. For ease of presentation, we again consider the most simple case where the mixing of the $F_i$ is 
dominantly with only one generation 
of the $f^a=f$ which, in the SM, obtains a mass from the usual Higgs mechanism, $m_f$. Such a possibility can be arranged, \eg, through the proper use of some 
flavor-dependent discrete symmetries as shown in Ref.\cite{Carone:2018eka}. Denoting this set of fermion fields in the original weak basis as 
${\cal F}^{0T}=(f^0,F^0_1,F^0_2)^T$, the general form of the various mass terms for these fermions can be written as $\bar {\cal F}^0_L{\cal M}{\cal F}^0_R +h.c.$ where 
\begin{eqnarray}
{\cal M} & =  \left( \begin{array}{ccc}
                         m_f & 0  & 0   \\
                          X_1 &  m_1 & 0 \\
                          X_2 & 0 & m_2 \\
                         \end{array}\right) \,,
\end{eqnarray}
which can be diagonalized, as usual, via a bi-unitary transformation $M_D=U_L{\cal M}U_R^\dagger$, where $M_D$ is diagonal so that ${\cal F}_{L,R}=U_{L,R}{\cal F}^0_{L,R}$ are 
the mass eigenstates. Again, as usual, $U_L$ is determined via the relation $M_D^2=U_L{\cal M}{\cal M}^\dagger U_L^\dagger$ while $U_R$ is similarly determined via 
$M_D^2=U_R{\cal M}^\dagger {\cal M}U_R^\dagger$. 

Before proceeding we need to say a few words about the values of the entries appearing in ${\cal M}$ (outside of the familiar $m_f$). As we will discuss in more detail below, the portal matter states 
$F_{1,2}$ can always be pair-produced at the LHC 
is the same manner as are the more conventional vector-like fermion that are usually discussed but which do not carry any dark charges. Thus we might semi-quantitatively expect that 
$m_{1,2}(T,B) \sim 1-2$ TeV while $m_{1,2}(E) \sim 0.2-0.5$ TeV or so~\cite{Egana-Ugrinovic:2018roi} 
based on the current null searches\cite{searches} at the LHC in which only SM particles are involved in the direct decay final state.  
(Of course, as we will stress below, these standard searches for vector-like fermion are not directly applicable in the present case under study.) On the other hand, since $v_s$ is at most a few GeV as  
we are interested in $m_V=g_Dv_s \lsim 1$ GeV or so, we might then expect that roughly $|X_{1,2}|\sim 1-$a few GeV unless the $\lambda_i$ are not O(1). 

With these typical values in mind we can straightforwardly obtain useful approximate analytic expressions for the elements of the matrices $U_{L,R}$ to lowest order in terms of the small 
set of four hierarchical ratios $|X_i|^2/m^2_j,~ i,j=(1,2)$ as 
\begin{eqnarray}
 U_L & =  \left( \begin{array}{ccc}
                         1-m^2_f[X^2_1/(2a^2_1)+X^2_2/(2a^2_2)] & m_fX_1^*/a_1  & m_fX_2^*/a_2   \\
                          -m_fX_1/a_1&  1-m^2_fX^2_1/(2a^2_1) & m^2_2X_1X_2^*/(\delta a_1) \\
                          -m_fX_2/a_2 & -m^2_1X_1^*X_2/(\delta a_2) & 1-m^2_fX^2_2/(2a^2_2)  \\
                         \end{array}\right) \,,
\end{eqnarray}
where we have defined $X^2_i=|X_i|^2$, $a_i=m^2_i-m^2_f$ and $\delta=m^2_1-m^2_2$; correspondingly, we also obtain
\begin{eqnarray}
 U_R & =  \left( \begin{array}{ccc}
                         1-[m^2_1X^2_1/(2a^2_1)+m^2_2X^2_2/(2a^2_2)] & m_1X_1^*/a_1  & m_2X_2^*/a_2   \\
                          -m_1X_1/a_1&  1-m^2_1X^2_1/(2a^2_1) & -m_1m_2X_1X_2^*/(\delta a_2) \\
                          -m_2X_2/a_2 & m_1m_2X_1^*X_2/(\delta a_1) & 1-m^2_2X^2_2/(2a^2_2)  \\
                         \end{array}\right) \,.
\end{eqnarray}
Here we see that, quite generally, the size of the elements in $U_R$ involving the mixing of the vector-like fermion with the SM fermion are significantly larger than the corresponding 
ones appearing in $U_L$. It is interesting to note that if we had instead chosen the portal matter fermions to be in $SU(2)_W$ doublets we would obtain the same results as above but with the 
roles of $U_L$ and $U_R$ simply interchanged. We note that in the corresponding case of the more `canonical' iso-singlet vector-like fermion that mix with the SM fields via the usual Higgs vev,  it 
is the $U_L$ matrix elements which are large and not the $U_R$ ones 
as is the case above. Here, since $U_L$ is close to the identity in the present case, as far as any mixing of the portal matter with the SM fields 
is concerned, this results in a further suppression of the possible portal matter decay into the familiar vector-like fermion final states involving the SM gauge bosons and Higgs field making these 
modes essentially irrelevant, vanishing to leading order in the $m_f\to 0$ limit. The three eigenvalues of $M_D^2$ are approximately given to leading order in the above mass ratios by 
$m^2_f(1+ \xi)$ and $m^2_i(1+\delta_i)$ with $\delta_i \simeq -X^2_i/a_i$ and 
\begin{equation}
 \xi \simeq -\frac{X^2_1a_2+X^2_2a_1}{a_1a_2} = \delta_1+\delta_2\,
\end{equation}
which demonstrates that all of the mass shifts due to mixing are generally quite small. Note that all of the expressions above simplify significantly for the cases where 
$f\neq t$ in which case the $m^2_f$ terms can always be safely neglected in comparison to the $m^2_i$. These mixings will then allow the portal matter to decay only to final states which are 
completely dominated by the presence of the dark fields $V,S$ and not the usual SM fields $W,Z,H$.

\section{Survey of Phenomenological Implications}

Our first goal is to understand how significantly the mixing of the SM fermion, $f$, with the vector-like fermion, $F_i$, alters the $f$ couplings to the dark photon, the SM and dark Higgs, as well as to the usual 
SM gauge bosons. Here we will work in the limit where $m^2_V/m^2_Z <<1$ so that in the original weak basis the part of the $V$ coupling to the SM due its KM with the hypercharge 
gauge boson is just $e\epsilon Q_f$ as noted above; note that such a coupling is now experienced by both the usual SM fields as well as by the portal matter ones since they also carry SM 
charges. Further note that in this limit of a small mass ratio, the dark photon does {\it not} pick up any additional couplings due to mass mixing with the $Z$ (\ie, the dark $Z$ scenario\cite{hoomy}) 
Suppressing Lorentz indices we may express this dark photon coupling in the notation above 
as $e\epsilon Q_f (\bar {\cal F}^0_L{\cal F}^0_L + L\to R) \to  e\epsilon Q_f (\bar {\cal F}_L{\cal F}_L + L\to R)$ as this part of the interaction is trivially insensitive to the mass 
mixing of $f$ with $F_{1,2}$. However, we recall that, unlike the SM fields, the portal matter themselves also have a direct coupling to $V$, $\sim g_D$, even when $\epsilon \to 0$; we can express 
this interaction in the form  $g_D (\bar {\cal F}^0_L C {\cal F}^0_L + L\to R)$ where $C$ is the just the diagonal matrix 
\begin{eqnarray}
C & =  \left( \begin{array}{ccc}
                         0 & 0  & 0   \\
                          0 &  1 & 0 \\
                          0& 0 & -1 \\
                         \end{array}\right) \,.
\end{eqnarray}
In the mass eigenstate basis this interaction then just becomes $g_D (\bar {\cal F}_L C_L {\cal F}_L + L\to R)$ where we define $C_{L,R}=U_{L,R}CU_{L,R}^\dagger$ with the 
elements of theses effective coupling matrices now given by $(C_R)_{ij}=U^R_{i2}U^{R*}_{j2}-U^R_{i3}U^{R*}_{j3}$ and similarly for $L\to R$ where $(i,j=1,2,3)$ corresponds to the 
mass eigenstates $f,F_1,F_2$, respectively. Some obvious but important things to note about these matrices include: ($i$) off-diagonal couplings to the dark photon between the SM $f$ and 
the vector-like fermion portal matter are generated allowing for decays such as $F_{1,2} \to fV$ with rates controlled by (the squares of) factors of order $m_iX_i/a_i$. ($ii$) $f$ picks up a mixing-induced 
{\it diagonal} coupling to the dark photon proportional to $g_D$ that is {\it not} vector-like, \ie, parity-violating, since $U_L\neq U_R$. Explicitly we find, using the matrices above that 
$(C_R)_{11}=m^2_1X^2_1/a^2_1-(1\to 2)$ while $(C_L)_{11}=m^2_fX^2_1/a^2_1-(1\to 2)$ and hence is smaller by factors of $m^2_f/m^2_{1,2}$ which are $<<1$ (except possibly in the 
case when $f=t$ but even there this factor is still likely quite small). We will return to the implications of these couplings in our discussion below. Note that whereas our single SM field 
picks up parity violating interactions with the dark photon by mixing with portal matter, in other scenarios\cite{hoomy} this happens via the mixing of the dark photon with the SM $Z$ to that {\it all} of the SM fields 
would have now have parity-violating interactions.

The mixing of $f$ with the portal matter can also lead to alterations in the couplings of these fields to the SM $W,Z$ bosons. In the $Z$ boson case, since $f_R, F_{1,2R}$ are all weak iso-singlets, 
the rotation of these fields by the matrix $U_R$ has no impact on their $Z$ couplings; this is no longer true for the corresponding left-handed fields. In the original weak basis these fields 
couple as $\frac{g}{c_W}(T_{3f}C_Z-x_WQ_f I)$ where $I$ is the $3\times 3$ identity matrix, $x_W=\sin^2 \theta_W$, $T_{3f}=\pm1/2$ is the usual third component of the 
weak isospin for the SM field $f_L$ and $C_Z$ is just
\begin{eqnarray}
C_Z & =  \left( \begin{array}{ccc}
                         1 & 0  & 0   \\
                          0 &  0 & 0 \\
                          0& 0 & 0 \\
                         \end{array}\right) \,.
\end{eqnarray}
The resulting elements of the left-handed coupling matrix in the mass eigenstate basis then become $\frac{g}{c_W}(T_{3f}U^L_{i1}U^{L*}_{j1}-x_WQ_f \delta_{ij})$ with $\delta_{ij}$ the 
usual Kronecker delta. There 
aportal matterre again several simple but important things to note: ($i$)  for the SM $f$, the quantity proportional to $T_{3f}$ now differs from unity by a term of order $m^2_fX^2_i/a^2_i <<1$ 
even for the top quark. Thus to a good approximation the couplings of the SM fields to the $Z$ are left unaltered by the mixing with portal matter; a similar argument can be made for the 
$W$ as well since the same factor of $U^L_{i1}U^{L*}_{j1}$ will also appear there. ($ii$) Off-diagonal couplings are generated such that decays such as $F_{1,2}\to fZ$ (as well as 
$f'W$) are now allowed but are controlled by (the squares of) factors of order $m_fX_i/a_i$. Note that these factors (squared) are smaller by $\sim m^2_f/m^2_{1,2}<<1$ than those 
controlling the decay $F_{1,2}\to fV$ described above; a similar suppression is observed in the case of the $W$. Hence the portal matter decays into the dark photon (and $S$ as we will find below) 
will {\it far dominate} those into the conventionally sought $W,Z$ final state modes which vanish in the $m_f\to 0$ limit. The impact of this will be discussed further in more detail below. 

Finally, we note that something similar happens when we consider the fermion couplings to the SM Higgs, $H${\footnote {We have set the $H-S$ mixing exactly to zero for this  
discussion due to the constraints mentioned above.}}. In the weak basis this is just $\frac{\sqrt 2 m_f}{v_H} \bar {\cal F}^0_L C_Z {\cal F}^0_R H+ h.c.$ using the notation above 
so that in the mass basis a coupling matrix, $C_H=U_LC_ZU_R^\dagger$, is produced whose elements are given by $(C_H)_{ij}=U^L_{i1}U^{R*}_{j1}$ and where we again see that 
the deviation of $(C_H)_{11}$ from unity is highly suppressed. Off-diagonal terms in $C_H$ allow for the decays $F_{1,2}\to fH$ but these are found to have suppressed (squared) 
couplings $\simeq m^2_fX^2_i/a^2_1$ similar to those for the $W,Z$ final states above so that these partial widths cannot remotely compete with those for $F_{1,2}\to fV$. This result 
should come as no surprise as this follows immediately from use of the Goldstone Boson Equivalence Theorem\cite{GBET}. In a similar manner, we can determine the corresponding coupling 
of the remaining real dark Higgs scalar, $S$, to the various fermions that are induced by the off-diagonal coupling above; we can express this result simply as 
$\beta_{ij} = [U^L_{i2}\lambda_1/\sqrt 2 +U^L_{i3}\lambda_2/\sqrt 2]U^{R*}_{j1}$.

\subsection{Leptonic Portal Matter}

We now turn to an examination of some sample implications of the above mixing and generated couplings for several choice of $f$; clearly the phenomenology will strongly depend 
upon the nature of the vector-like fermion portal matter 
and what SM field(s) it mixes with via the dark Higgs. Here we will limit ourselves to a few simple examples, reminding ourselves that in a more complete 
theory the situation may be much more complex than what we consider here.

\begin{figure}[htbp]
\centerline{\includegraphics[width=5.0in,angle=0]{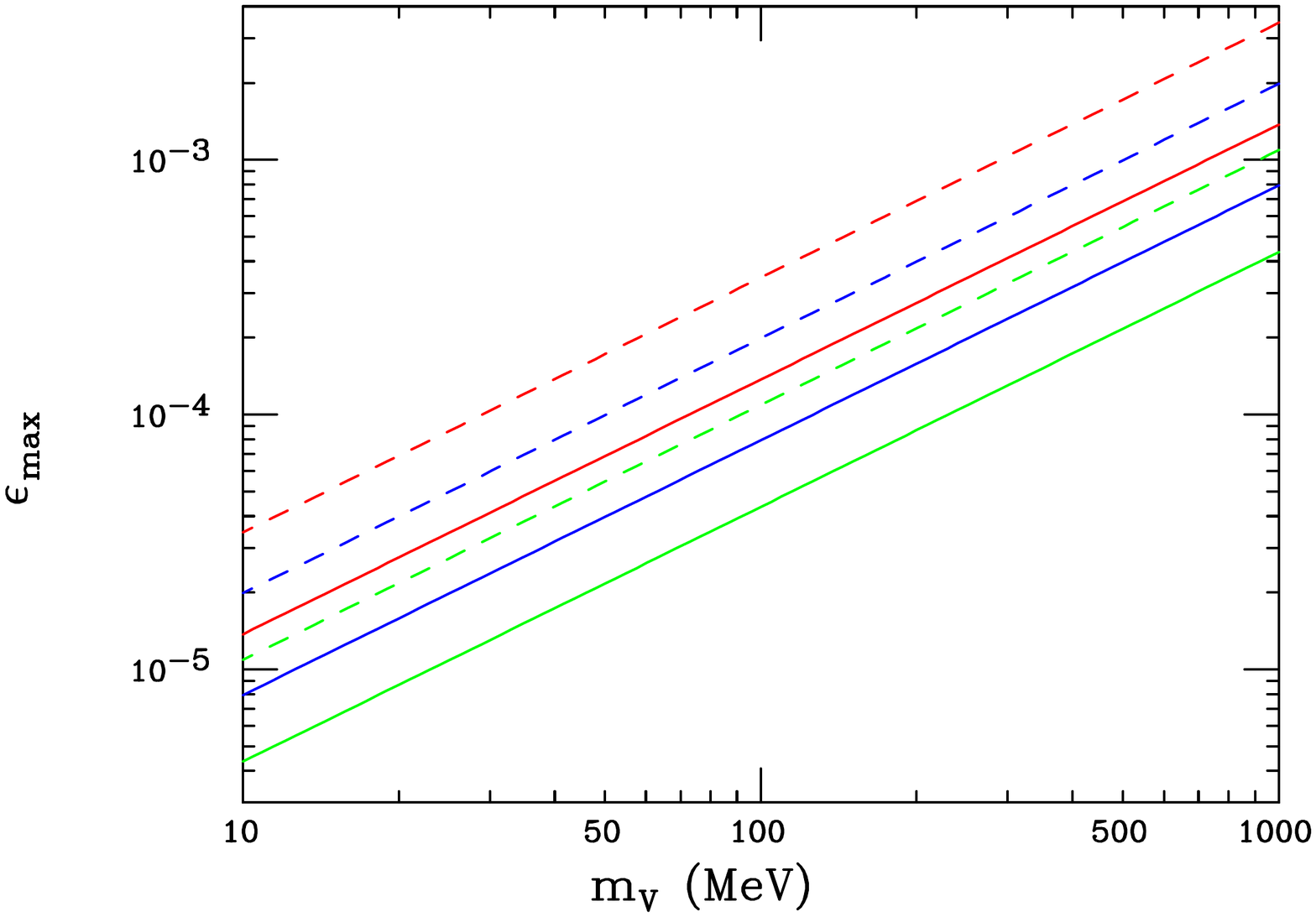}}
\vspace*{-2.5cm}
\centerline{\includegraphics[width=5.0in,angle=0]{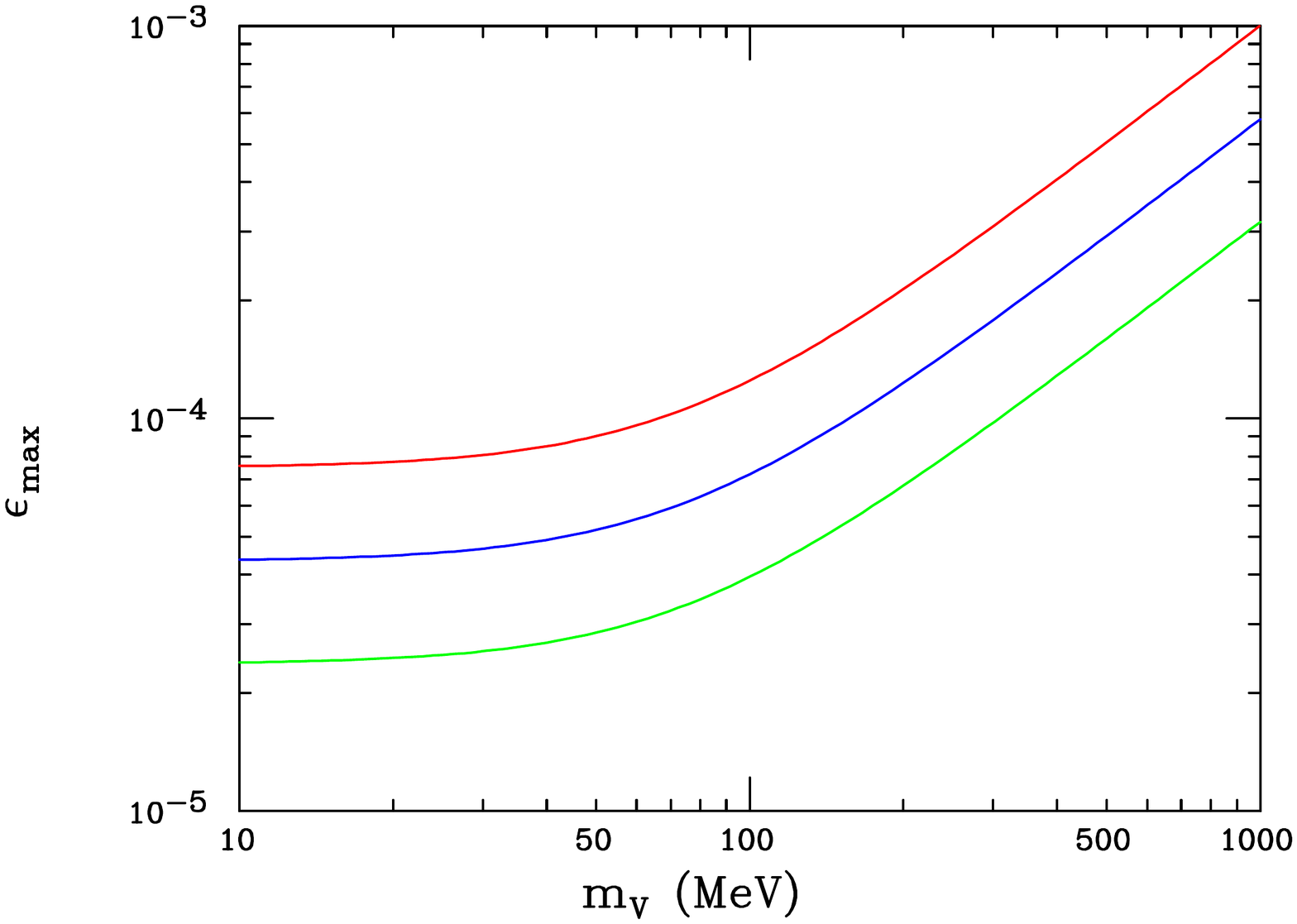}}
\vspace*{-1.50cm}
\caption{(Top) Maximum value of $\epsilon$ as a function of $m_V$ allowed by the present APV data for positive (negative) values of $y$ as solid (dashed) lines. From top to bottom,  
$|y|=0.1,0.3$ or 1 has been assumed, respectively. 
(Bottom) Parameter space to be probed by the MOLLER experiment assuming the SM result is recovered. Here, again we see the maximum value of $\epsilon$ as a function of $m_V$ for,
from top to bottom $|y|=0.1,0.3,q.0$, respectively.}
\label{ref-plot}
\end{figure}

We first consider the case of a vector-like lepton, \ie, $F=E$, 
so that $f=e$ above; in this case $U_L \simeq I$ (at least as far as the $ee$ and $eE_{1,2}$ elements are concerned) due to the small value of $m_e$ relative to all the relevant mass 
scales and so we can completely neglect left-handed mixing as far as SM fields are concerned in 
the rest of the present discussion. However, the right-handed mixing is still active so that, collecting the various pieces, \eg, the coupling of the electron to the dark photon is now given by 
\begin{equation}
\big[-e\epsilon ~\bar e\gamma_\mu e +g_D(C_R)_{11}~\bar e\gamma_\mu P_R e\big]V^\mu \equiv -e\epsilon ~\bar e \gamma_\mu (v_e-a_e\gamma_5)eV^\mu\,,
\end{equation}
where $v_e=1-y$ and $a_e=-y$ with $y=g_D(C_R)_{11}/2e\epsilon$. One might roughly expect that $(C_R)_{11}\simeq X^2_i/m^2_i \simeq 10^{-4}$ in the present case based on the 
arguments above assuming that $m_E\sim 2-300$ GeV, so that, very roughly,  $0.01 \lsim |y| \lsim 0.5$ and thus the dark photon may now mediate a reasonably strong 
parity-violating interaction for the electron. This possibility may be probed~\cite{hoomy} by a 
number of low-energy (\ie, low momentum transfer, $Q^2 \lsim m^2_V$) experiments  such as in atomic parity violation (APV)\cite{APV} or in polarized electron-electron/electron-proton 
scattering\cite{MOLLER}. In the case of APV, the amount of parity violation is quantified by a parameter $Q_W$, the weak charge, which in the SM is just 
$-(A-Z)+(1-4x_W)Z \simeq -73.23(1)$\cite{PDG} for the case of Cesium ($Z=55, A=133$) whereas the experimental value is -72.62(43)\cite{PDG}. dark photon exchange in the present setup 
produces an additional contribution to $Q_W$ given by: 
\begin{equation}
\Delta Q^{V}_W= \frac{-8{\sqrt 2}\pi \alpha \epsilon^2}{G_Fm^2_V} y Z \frac{m^2_V}{Q^2+m^2_V}\,,
\end{equation}
where $G_F$ is the Fermi constant and which is given numerically (with $Q^2 \to 0$) by
\begin{equation}
\Delta Q^{V}_W(Cs)\simeq -1.22~y ~\Big(\frac{100 \rm MeV}{m_V}\Big)^2 \Big(\frac{\epsilon}{10^{-4}}\Big)^2\,,
\end{equation}
which must satisfy the constraint $-0.23\leq \Delta Q^{V}_W\leq 1.45$ at $95\%$ CL to be consistent with the current experimental result. This is relatively easily done for values of 
$|y|$ in the above range of interest even for correspondingly respectable ranges in the values of both the parameters $m_V$ and $\epsilon$. The parameter space probed here is 
shown in the top panel of Fig.~\ref{ref-plot}.

Polarized Moeller scattering\cite{E158,MOLLER},  
on the other hand, will offer a different probe of $y \neq 0$ as it leads to an apparent shift in the effective value of the weak mixing angle which can be written as 
\begin{equation}
\Delta x_W \simeq  5.56 \cdot 10^{-3} ~y ~\Big(\frac{100 \rm MeV}{m_V}\Big)^2 \Big(\frac{\epsilon}{10^{-4}}\Big)^2 \frac{m^2_V}{<Q^2>+m^2_V}\,,
\end{equation}
with $<Q^2>$ being effective average value of $Q^2$ probed by the experiment which in this case is $\simeq (75 ~\rm MeV)^2$\cite {APV,MOLLER}. Assuming that the SM value is 
realized at the end of the MOLLER experiment, this will result in a constraint  $|\Delta x_W| \leq 5.6\cdot 10^{-4}$. For $m_V=100(300)$ MeV with $\epsilon=10^{-4}$ this will imply a 
bound of $|y|\leq 0.16(1.2)$ which we again see can be easily satisfied. Quite generally, however, this experiment is observed to be able to probe a reasonable portion of the parameter 
space of this particular setup; of course a result in conflict with SM expectations would prove most interesting. The parameter space probed by MOLLER is 
shown in the lower panel of Fig.~\ref{ref-plot} assuming that the SM expectation is obtained; comparison with the APV show that these two probes are complementary. 

\begin{figure}[htbp]
\centerline{\includegraphics[width=5.0in,angle=0]{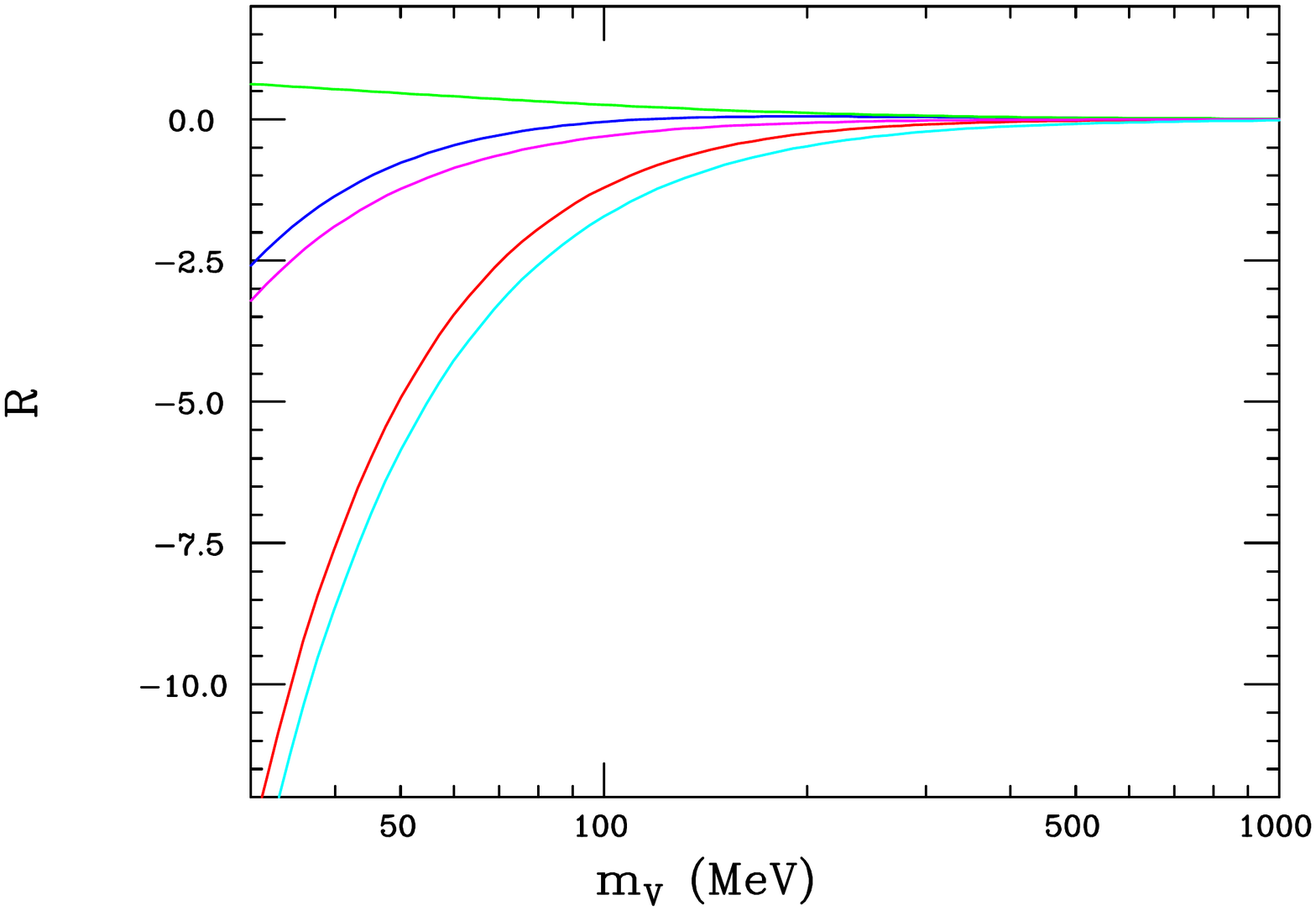}}
\vspace*{-1.50cm}
\caption{
The parameter $R$, as defined in the text, as a function of the $V$ mass $m_V$ assuming $y=-0.5(-0.25, 0,0.25.0.5)$ corresponding to the red(blue, green, magenta, cyan) curves, 
respectively. 
}
\label{gm2}
\end{figure}

In the case where portal matter is muon-like instead of electron-like, \ie, $F=M$, one might ask if the added flexibility allowed by $y\neq 0$ could help explain the anomaly\cite{Jegerlehner:2018zrj} 
in the muon's value of g-2\cite{PDG}; this question is easily addressed employing the general results given in\cite{Leveille:1977rc}. There are actually several new contributions at the 
1-loop level to this quantity 
beyond the ones usually appearing in dark photon models\cite{hoomy,Davoudiasl:2012ig}. Ordinarily in such models the lone contributing diagram has a dark photon loop with the SM photon emitted 
off the internal muon line and where the dark photon coupling to the muon is both 
diagonal and purely vectorial; such a graph is usually simply suppressed by the small value of $\epsilon^2$~\cite{hoomy}. In the present scenario such a diagram exists but both vector and 
axial-vector couplings are now present. Generally the axial-vector contribution is found to add destructively with the one arising from the vector coupling making it less likely to 
explain the observation. Furthermore, this axial-vector contribution tends to be the numerically 
larger one so that the sum of both contributions is {\it negative} yielding the wrong sign from what is needed 
to explain the deviation from the SM. Employing the notation above we find the sum of these two contributions can be expressed as 
\begin{equation}
\Delta g^{V_1}_\mu= 2.32\cdot 10^{-11} \Big(\frac{\epsilon}{10^{-4}}\Big)^2 \int_0^1dx~\frac{x^2(1-x)(1-y)^2-x(1-x)(4-x)y^2-2x^3y^2r^{-1}}{x^2+r(1-x)}\end{equation}
where $r=m^2_V/m^2_\mu \sim 1$ and explicitly shows the opposite sign contribution arising from the axial-vector coupling term. The magnitude of thus 
integral is usually $\lsim $ a few so that this overall 
contribution is quite small for our default parameter choices; for example, taking $r=1$ or larger one generally obtains rather small values for the integral and which 
can be of either sign depending upon the specific value of $y$. To obtain a more general understanding of this result, we can rewrite the expression above as 
\begin{equation}
\Delta g^{V_1}_\mu= 10^{-11} \Big(\frac{\epsilon}{10^{-4}}\Big)^2 R(y,m_V)\,,
\end{equation}
where the quantity $R$ is shown in Fig.~\ref{gm2} as a function of $m_V$ for various values of $y$. Clearly, for our parameter choices 
$R$ is almost always negative and generally only yields at most a small overall contribution to the g-2 of the muon. 

In addition to these `standard' dark photon contributions, there are now also a pair of 1-loop graphs with either $V$ or a dark Higgs loop\cite{Chen:2015vqy} where the internal line is a portal matter  
state, $M_i$, and the couplings at both vertices are chiral. We note that in the conventional dark photon model, a dark Higgs contribution usually arises due to its mixing with the SM Higgs 
and the muon resides on the internal line\cite{Chen:2015vqy}. Here we find that the dominant dark Higgs contribution instead arises from the muon's mixing with the analogous portal 
matterstate with 
the portal matter appearing on the internal fermion line. For these types of contributions the strong suppression is now seen to be the result of the rather small value of the ratio 
$m^2_\mu/m^2_i$  for the typical values of the portal matter masses that we might expect. In the scalar case we find (since $m^2_S<<m^2_i$) that a positive result is obtained, \ie, 
\begin{equation}
\Delta g^S_\mu= 5.85\cdot 10^{-10} \sum_i \frac{X^2_i}{v^2_s} \Big(\frac{200 \rm GeV}{m_i}\Big)^2\,,
\end{equation}
which, since $|X_i|/v_s \sim 1/2$, is roughly an order of magnitude too small to make an important contribution unless the values of $m_i$ are significantly smaller than we might 
naively expect\cite{Egana-Ugrinovic:2018roi} . For the corresponding $V$ graph with portal matter on the internal line, we instead obtain
\begin{equation}
\Delta g^{V_2}_\mu =-2.35\cdot 10^{-4} ~\Big(\frac{g_D}{0.1}\Big)^2~\Big(\frac{100 \rm MeV}{m_V}\Big)^2 \sum_i \Big(\frac{X^2_i}{m^2_i}\Big)\,,
\end{equation}
where we have used the fact that $m^2_i>>m^2_{V,\mu}$ and that $|(C_R)_{1i}| \simeq |X_i|/m_i$ to leading order in the small mass ratios. Given the estimates above this 
contribution will also yield a {\it negative} result but only as most as large as O($10^{-11}$), similar to the first dark photon contribution obtained above. 

Finally, we stress our overall result that in this current setup, $g-2$ does not constrain any of the model parameter space of interest to us particularly due to the multiple parameters that 
enter into the calculations and the corresponding numerical freedom that they represent.

It is interesting to now very briefly consider how this singlet electron- or muon-like portal matter case differs from an alternative setup where the portal matter fields form the vector-like 
lepton  $SU(2)_W$ doublet $(N,E)^T$ (or $(N,M)^T$). In 
the $Q=-1$ sector, all of the analysis presented in the previous discussion, with the interchange of left-handed and right-handed labels, essentially remains unaltered except 
that $y\to -y$. The 
presence of the $Q=0$, $N$, however, brings us something new. First, note that due to $SU(2)_W$ the weak states $N_i$ and $E_i$ will have essentially the same masses 
so that the $U_L$ matrices for both the $Q=0,-1$ sectors will be the same. This implies, in the limit that lepton masses can be neglected, that there is no new physics in the charged 
current sector coupling to $W^\pm$. However, the mixing of $N$ with the ordinary left-handed 
SM $\nu$ field now induces a coupling of this SM $\nu$ field to the dark photon\cite{Bilmis:2015lja} of the form
\begin{equation}
2e\epsilon y~(\bar \nu \gamma_\mu P_L \nu) ~V^\mu\,,
\end{equation}
which may lead to observable effects in low energy neutrino experiments provided that the typical $<Q^2>$ is not very much larger than $m^2_V$\cite{nuints}. We note in passing 
that when finite $m^2_V/m^2_Z$ corrections are included in the usual KM model (where portal matter effects are not considered), couplings to the neutrino are also induced as well but then 
these have a flavor universal strength suppressed by this very small mass ratio in comparison to $e\epsilon Q${\footnote {This possibility was also briefly discussed 
in\cite{Davoudiasl:2012ig}.}} 
These couplings, where all of the neutrinos would be effected as opposed to a single flavor in the present case, are far smaller than the ones that can potentially be induced by 
mixing of the SM fields with the portal matter since $|y|$ can be sizable as discussed above. This possibility will be explored elsewhere\cite{tomorrow}.

\subsection{Hadronic Portal Matter}

Now let us turn our attention to our main emphasis, \ie, the perhaps more interesting 
case where the portal matter fields also carry color which to our knowledge has never been considered previously. Here we mainly 
focus on the specific example of the bottom-like singlet vector-like quark, \ie, $F_i=B_i$, acting as the portal matter. While the case 
$F_i=T_i$ shares several features with this possibility, it has many unique aspects of its own which we will briefly discuss and contrast with the $F_i=B_i$ scenario below. 
For $F_i=B_i$, the general mixing/coupling analysis presented above remains applicable as $m^2_b<<m^2_{1,2}$. Instead of flavor physics and 
low-energy phenomena, here we will be more interested in $B_i$ production and decay at the LHC especially as we anticipate that roughly $m_{1,2}\sim 1-2$ TeV in this 
case{\footnote {Of course, if there were additional flavor symmetries active in the quark sector, the implications on low energy phenomenology of such new states that might be involved 
in rare $B$ decays could also be very interesting.}}. 

\begin{figure}[htbp]
\centerline{\includegraphics[width=5.0in,angle=0]{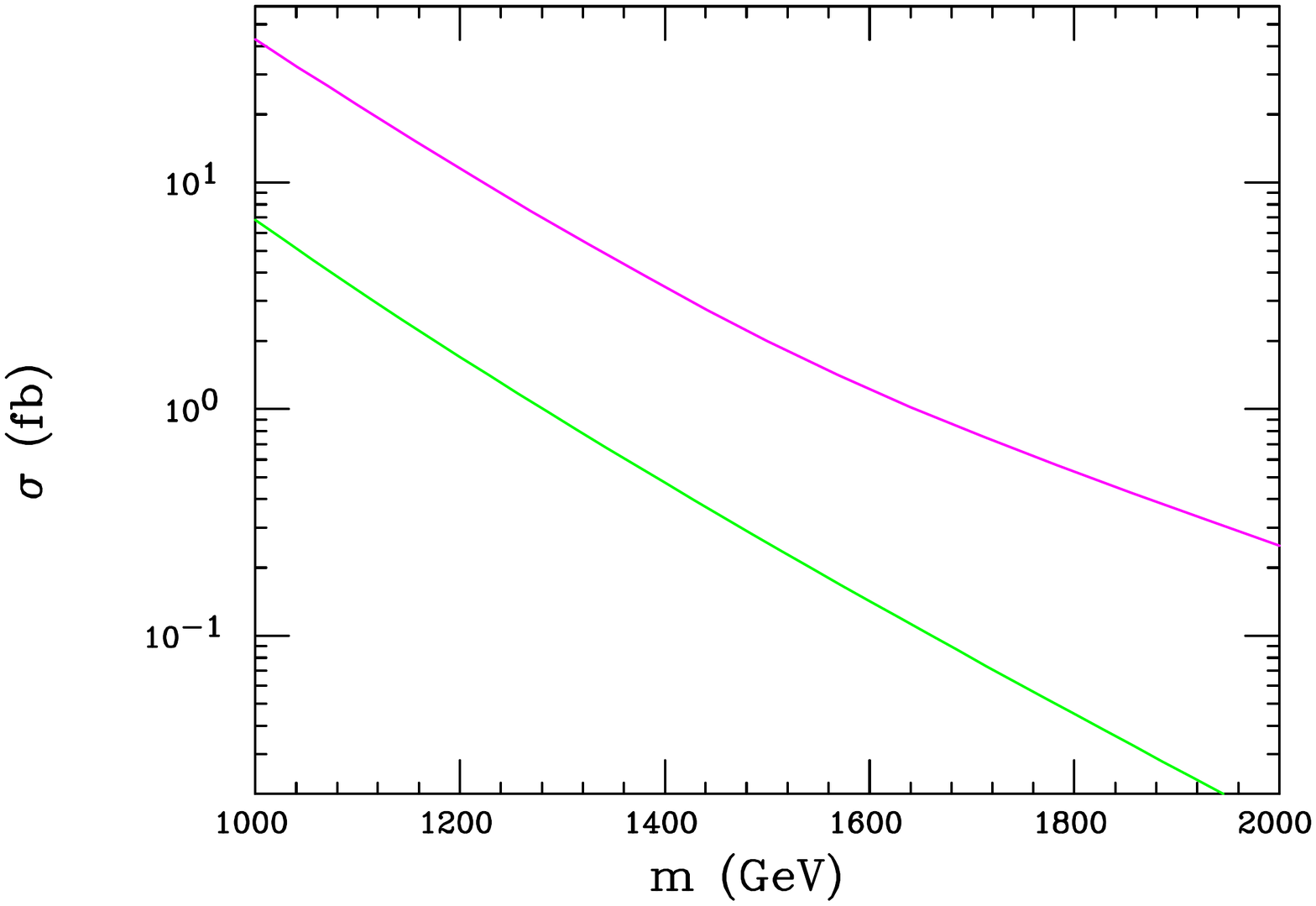}}
\vspace*{-2.5cm}
\centerline{\includegraphics[width=5.0in,angle=0]{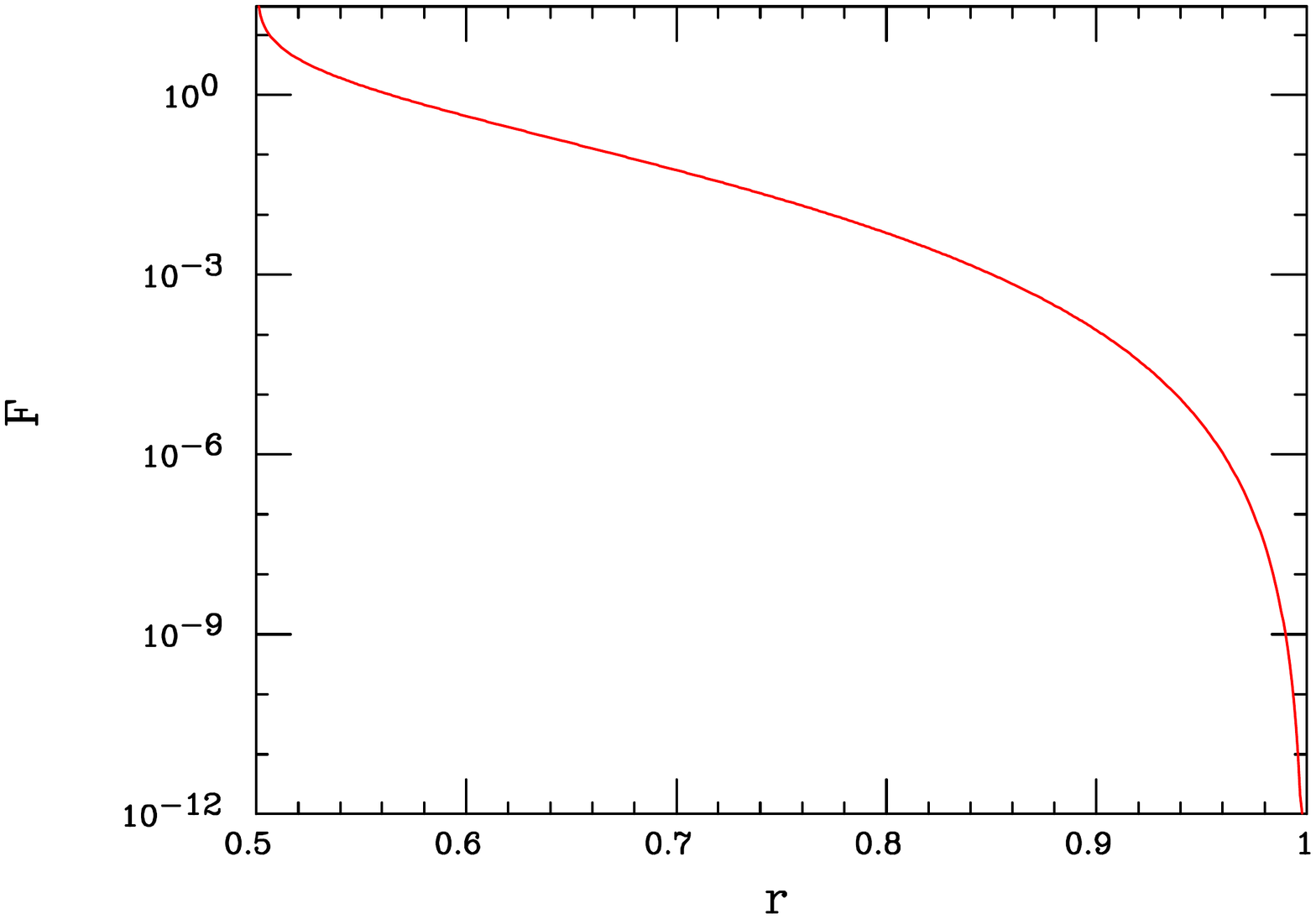}}
\vspace*{-1.50cm}
\caption{(Top) $B\bar B$ (magenta) and b-squark (green) pair production cross section at $\sqrt s=13$ TeV as discussed in the text.
(Bottom) Values of the 3-body phase space function $F$ for the decay $S\to Ve^+e^-$ as a function of $r=m_V/m_S$ as described in the text.}
\label{svstar}
\end{figure}

We first examine the standard QCD 
pair-production mechanism which is model-independent except for the values of $m_{1,2}$. To set the scale for the overall event rate, note that at $\sqrt s=13$ TeV the production cross 
section for $\bar BB$ production at the LHC would be roughly $\simeq 42(2.0, 0.60, 0.25)$ fb for $m_B=1(1.5,1.8,2)$ TeV\cite{production}, respectively, as shown in the top 
panel of Fig.~\ref{svstar}. 
Generally, for this interesting mass range, this cross section is seen to scale roughly as $\sim m_B^{-7}$ so that values for other nearby masses can be easily obtained.  This also 
implies that if $m_{B_2}/m_{B_1}=1.5(1.8)$ then $\sigma(B_2)/\sigma(B_1)\lsim 0.06(0.02)$ so that almost the entire signal would arise from the lighter of the two states; we will 
mostly concentrate our analysis on the lighter of these two states below. As noted 
above (and as now applied specifically to this example), one easily finds that the partial decay widths for the conventional modes such as $B_i\to bZ,bH$ and $B\to tW$ to be quite 
highly suppressed in comparison to the partial widths for decays to the $bV,bS$ final states, vanishing completely in lowest order in the $m_b\to 0$ limit. This implies that the ordinary 
modes usually employed at the LHC in the searches for vector-like quarks are {\it inoperable} in this case as noted previously. 

How do the partial widths for the two $B_i\to bV,bS$ decay modes compare? We first recall that both the $B_i$ are likely to have qualitatively comparable masses such that they satisfy 
$m^2_{1,2}>>m^2_b, m^2_{S,V}$ since both $m_{V,S}$ are $\sim$ a few GeV at most. In the limit where we can neglect all of the light particle masses appearing in the final state we find 
that the partial widths for both the $V,S$ modes are identical (again this is due to the Equivalence Theorem\cite{GBET}) and are given in this small mixing angle, small mass-squared 
ratio limit by
\begin{equation}
\Gamma(B_i \to bV,bS)=\frac{\lambda^2_i m_i}{32\pi}\,,
\end{equation}
where the $\lambda_i$ have been defined above and are nominally of O(1). These decays are clearly prompt even for substantially smaller values of the $\lambda_i$'s. It is 
important to remind ourselves of the lifetimes of the $S,V$ states that appear here as $B_i$ decay products. (The decay of the heavier $B_2$ state into the lighter $B_1$ plus $V$ 
or $S$ is found to be suppressed in rate by factors of order $\sim v^2_s/m^2_2$\cite{tomorrow} in comparison to that to the $bV,S$ final state. ) The well-known total 
decay width\cite{KM,vectorportal} for $V$ can 
be written in terms of its partial width into the $e^+e^-$ final state and the branching fraction for this mode, $B_e(m_V)$, as (taking here the $m^2_e/m^2_V \to 0$ limit for 
simplicity given the $V$ mass range of interest below)
\begin{equation}
\Gamma(V\to all)=\frac{\alpha \epsilon^2 m_V}{3B_e}= 2.432\cdot 10^{-9}~\Big(\frac{\epsilon}{10^{-4}}\Big)^2 \Big(\frac{m_V}{100 \rm MeV}\Big) \frac{1}{B_e} ~\rm MeV\,,
\end{equation}
which corresponds to an {\it un-boosted} decay length for $V$ of 
\begin{equation}
c\tau \simeq 81.44~\Big(\frac{10^{-4}}{\epsilon}\Big)^2 \Big(\frac{100 ~\rm MeV}{m_V}\Big) B_e ~\mu m\,.
\end{equation}
Note that for the corresponding $V$ decay to pairs of muons above threshold, the above electronic partial width must be scaled by a factor of $\beta_\mu (3-\beta^2_\mu)/2$ where 
$\beta^2_\mu =1-4 m^2_\mu/m^2_V$. In our numerical analyses, in order to obtain the relevant value of $B_e$, we have employ the $e^+e^- \to hadrons$ results and corresponding 
treatment as described in Ref.~\cite{sample}.  

Now recall that $m_V/m_S=g_D/\sqrt{2\lambda_S}$ so it is very likely that $m_S>m_V$ and we will assume that this is true in what follows. When $m_S>2m_V$, then the dominant 
$S$ decay is naturally $S\to VV$ via the vev, $v_s$, with a partial width given by{\footnote {Recall that in the present setup, $S$ has no tree-level allowed decay to DM assuming that such 
a mode is even kinematically open.}}
\begin{equation}
\Gamma(S\to VV)=\frac{g^2_D m_S}{128\pi} \frac{1}{x_V} (1-4x_V)^{1/2} (1-4x_V+12x^2_V)\,,
\end{equation}
where $x_V=m^2_V/m^2_S$ and this is also seen to be a prompt decay, even near threshold. If instead we have that $m_V<m_S<2m_V$, we obtain a partial width of 
\begin{equation}
\Gamma(S\to Ve^+e^-)=\frac{g^2_D\alpha \epsilon^2 m_S}{96\pi^2}~F(r)= 7.7\cdot 10^{-16} ~m_S\Big(\frac{g_D}{0.1}\Big)^2 \Big(\frac{\epsilon}{10^{-4}}\Big)^2F(r)\,,
\end{equation}
where here $1/2 <r=m_V/m_S=\sqrt x_V <1$ and $F$ is a well-known function\cite{longago} that ranges over many orders of magnitude and whose behavior we display in 
the lower panel of Fig.~\ref{svstar}.  We note, for example, that for $r=0.6(0.7,0.8,0.9)$ one finds that $F=0.443(5.55\cdot 10^{-2}, 4.97\cdot 10^{-3}, 1.18\cdot 10^{-4})$, 
respectively, so that $F$ falls off quite rapidly with increasing $r$, and this can lead to a very substantial width suppression/lifetime lengthening due to phase-space. Of course, very close 
to the $VV$ threshold finite $V$ width effects will become important here but as noted above the $V$ total width itself is quite small due to the appearance of the overall $\epsilon^2$ 
suppression factor. Since it is clear that this $S\to Ve^+e^-$ partial width is 
extremely small we must take some care that $S$ has no other possible decay modes than may compete, \eg, through the very suppressed mixing with the SM Higgs via 
a tiny, but non-zero value of $\lambda_{HS}$ as discussed above.  For example, the $H-S$ mixing-induced partial width for $S\to l^+l^-$ is given by 
\begin{equation}
\Gamma(S\to l^+l^-)\simeq \frac{G_F m^2_l m_S}{4\sqrt 2 \pi} ~\beta^3_l \theta^2_{HS}\lsim 2.4\cdot 10^{-16}m_S\,,
\end{equation}
where in the last step we consider the case of $l=\mu$, have set the phase space factor $\beta_l\to 1$, employed the upper limit on the value of $\theta_{HS}$ discussed above with 
$v_s/v_H=10^{-2}$ and assumed an invisible Higgs branching fraction of $B_{inv}=0.1$. (The corresponding partial width into $e^+e^-$ is smaller by roughly a factor of 
$\sim 4.3 \cdot 10^4$.) Here we see that if $S$ lies above the $\mu^+\mu^-$ threshold and we saturate the 
LHC bound on the $H-S$ mixing coming from SM Higgs decay data then the rate for the $S\to \mu^+\mu^-$ decay may be comparable to or even dominate over that for $Ve^+e^-$ especially 
since the function $F$ is potentially very small. Here we will {\it assume} that this is {\it not} the case, \eg, we will always assume that the mixing of $S$ with the SM Higgs is so small that it 
has no influence upon how $S$ decays; we proceed with this caveat kept in mind.
Inserting typical values into the above expression for the $S\to Ve^+e^-$ partial width leads to an {\it un-boosted} $S$ decay length for masses in the relevant interval of  
\begin{equation}
c\tau \simeq 170.8 ~\Big(\frac{0.1}{g_D}\Big)^2\Big(\frac{10^{-4}}{\epsilon}\Big)^2 \Big(\frac{150 ~\rm MeV}{m_S}\Big) \frac{1}{F(r)} ~\rm cm\,.
\end{equation}
We note that {\it extremely} long proper lifetimes for $S$, of order ~1 s or more, which could result in some parameter space regions, are likely excluded by cosmological considerations 
and particularly by the bounds that arise from nucleosythesis constraints\cite{Ruchayskiy:2012si}.

\begin{figure}[htbp]
\centerline{\includegraphics[width=5.0in,angle=0]{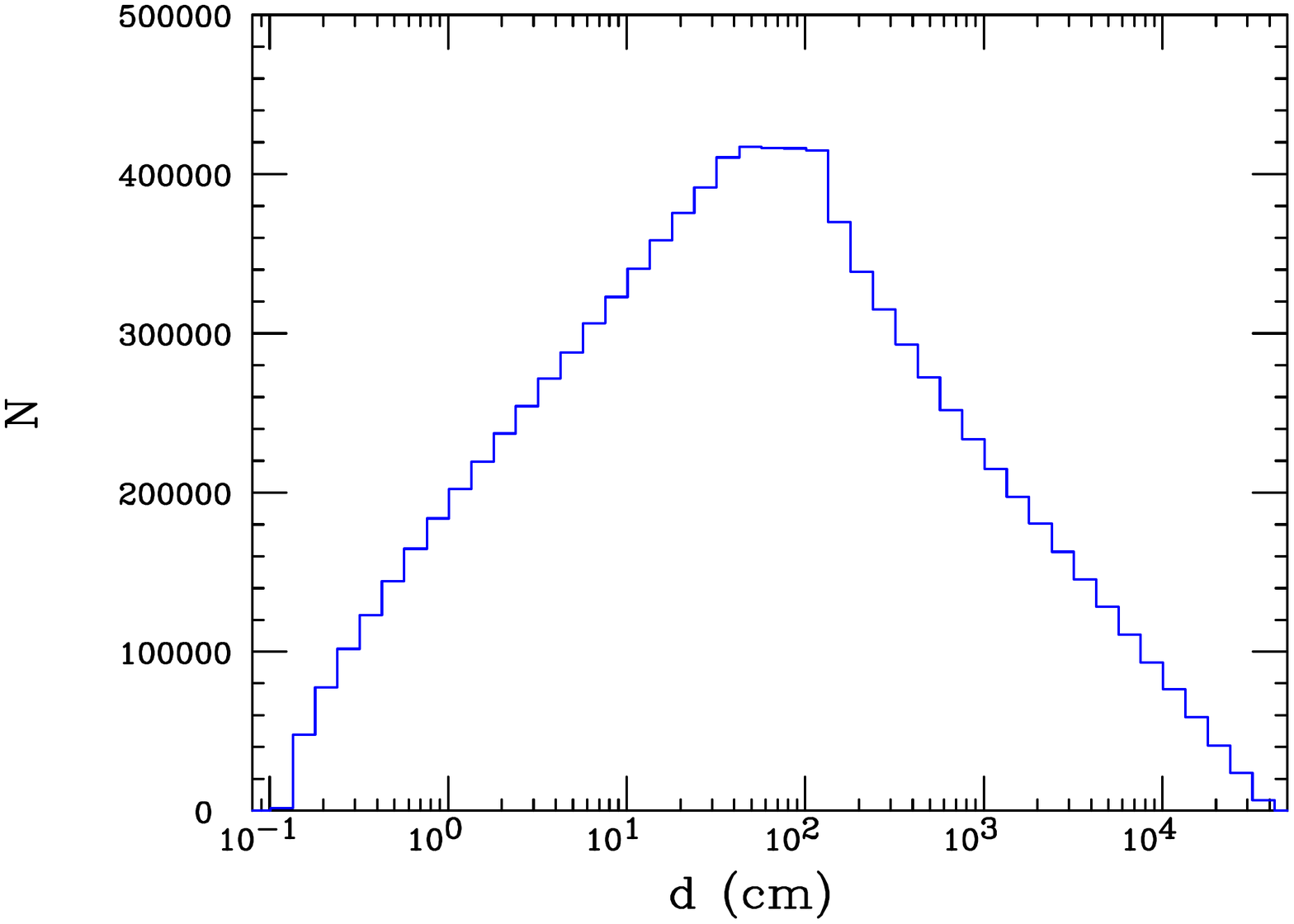}}
\vspace*{-2.5cm}
\centerline{\includegraphics[width=5.0in,angle=0]{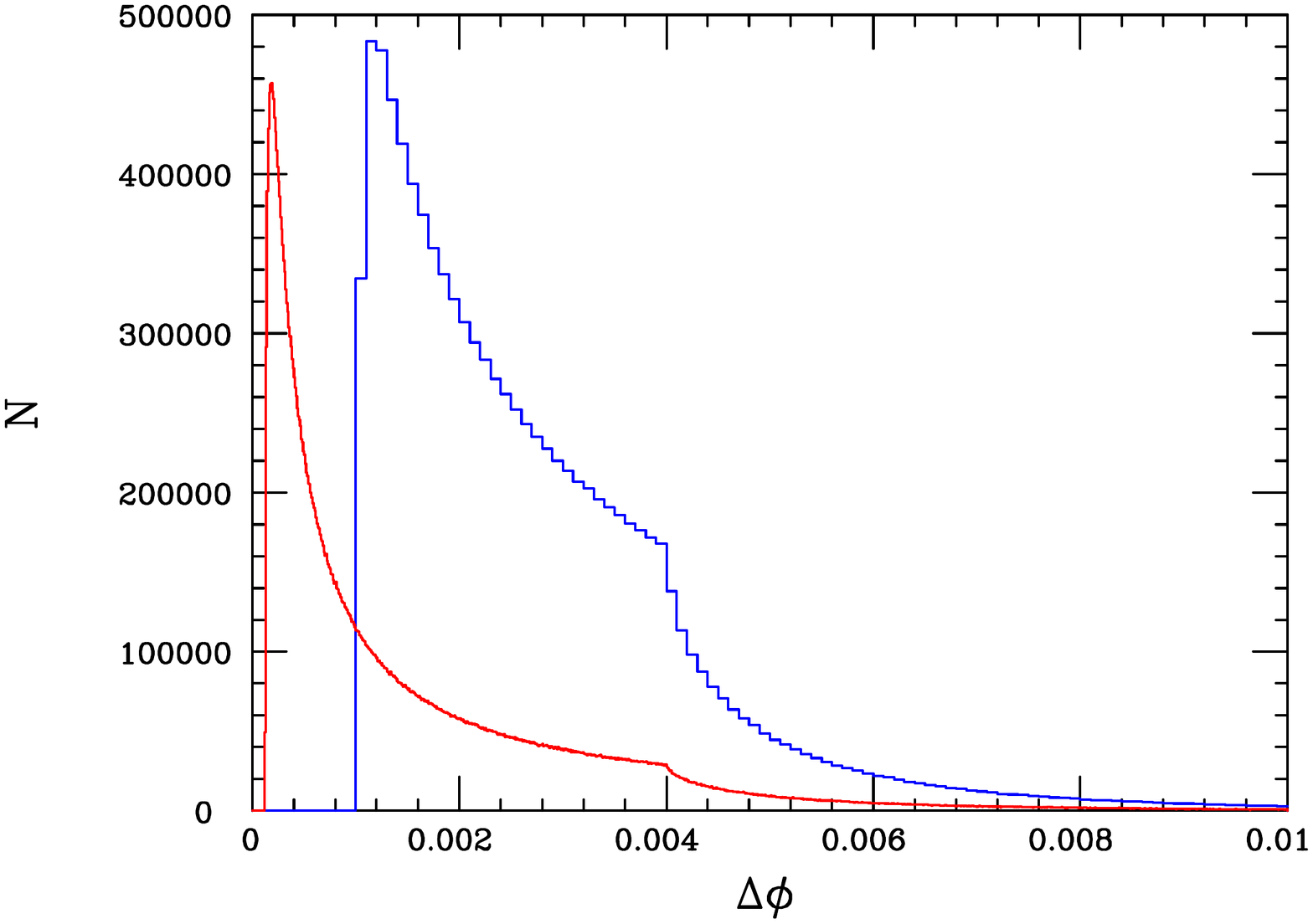}}
\vspace*{-1.50cm}
\caption{
(Top) Boosted decay length ($d$) distribution of the dark photon from the decay $B\to bV$ assuming that $m_B=1$ TeV as described in the text.
(Bottom) Boosted leptonic opening angle distributions, averaged over the model parameter space, for the boosted $V\to e^+e^-$ (red)  and $V\to \mu^+\mu^-$ (blue) decay modes.}
\label{vlength}
\end{figure}

From these estimates it would seem that bottom-like portal matter (generically $B$ for simplicity) pair-production at the LHC would produce quite unusual events in certain phase space 
regions, \ie, the apparent signatures for $B$ pair production depend critically upon where one sits in the model parameter space. Perhaps the simplest situation 
is where both the $B$ and $\bar B$ each promptly decay into a dark photon, \ie, the $b\bar b VV$ final state.  Since the $B$'s are presumed heavy and relatively slow their decay will yield two, 
high $p_T$, b-jets (that can act as the trigger) which are generally {\it not} back to back. Both $V$'s will be very highly boosted and also will likely not be back to back, each sharing 
the parent $B$ rest energy with the associated $b$-jet, so that the resulting boost is roughly given by 
$\gamma_V \simeq m_B/2m_V \sim 5000(1000)$ if $m_V=100(500)$ MeV and $m_B=1$ TeV; the boost is, of course, somewhat larger if $B$ is 
even more massive. To get a rough idea of the resulting decay properties and {\it boosted} decay lengths ($d=\gamma_V \beta c\tau$) of the dark photon within the interesting parameter 
space region we perform the following scan: ($i$) we take $\epsilon$ 
to lie in the range $10^{-5}-10^{-3.5}$ while, independently, ($ii$)  $m_V$ to lie in the range $10^{1.5}-10^3$ MeV.  Specific values in these ranges are chosen employing log priors. Once 
$m_V$ is known we can then estimate the value of $B_e(m_V)$\cite{sample} and hence determine the quantity of interest, $d$, using the equation above. We note that the low-mass end of 
this range where $\epsilon$ is simultaneously chosen to also be at the lower end of its range is somewhat disfavored by existing experiments\cite{Battaglieri:2017aum}. We 
do {\it not} account for this as part of the  
scan but point out that it is in this region that the longest $V$ lifetimes (and corresponding large boosts) are frequently obtained. Given these numerical assumptions and caveats, the 
result of this scan for $10^7$ generated model points is shown as the histogram in the top panel of Fig.~\ref{vlength}. For these assumed parameter ranges, the peak of this distribution is 
seen to lie roughly at $d\sim 0.7$ m but the distribution of possible values for $d$ is observed to have quite long, slightly asymmetric tails going down to values of $d\sim 2$ mm and up to 
$d\sim 400$ m. In all cases $m_B=1$ TeV has been assumed here; larger mass values will lead to correspondingly increased boosts but only by a small factor. Of course, any given 
model point 
in this parameter space will realize only a {\it specific} $d$ value within this range. We note the trivial observation that for a fixed value of $d$, the two $V$'s in the final state can generally 
decay at different distances from the IP and these may even occur in different LHC detector elements.

For those regions of the parameter space where {\it both} $V$ decays will very likely occur long after they leave the detector, such events would essentially appear (incorrectly) as having 
missing $E_T$ (MET) in addition to the two $b$-jets. This is a `conventional' signature that is essentially covered by the usual SUSY direct b-squark decay to bottom plus the LSP 
searches\cite{bsquark}, albeit with somewhat different production cross sections and acceptances{\footnote {Note that the total b-squark pair production cross section in this mass 
range for decoupled gluinos is roughly a factor of $\sim 6-10$ smaller than is that for $B\bar B$ pairs\cite{bsig,Beenakker:2016lwe} as shown in the upper panel of Fig.~\ref{svstar}.}}.  
The present null results for these bsquark searches in the limit of a massless neutralino (which most closely corresponds to the present possibility under discussion) certainly 
would imply that $m_{1,2}>1-1.5$ 
TeV in such cases. If we had instead considered the top-like portal matter, $T$, which would similarly decay to a $tV$ final state with a very long lived $V$, such signatures would be then be 
covered by the analogous top squark searches leading to somewhat similar, but slightly weaker, mass constraints on the portal matter due to the reduced efficiency of top reconstruction. 
Of course, if a secondary, `far' detector were to be available at an appropriate distance, such as MATHUSALA\cite{Alpigiani:2018fgd} or FASER\cite{Ariga:2018zuc}, 
a respectable fraction of these long-lived particle decays 
might be captured and studied. 

Perhaps more interesting are events where one or both of the $V$'s decay before/inside the calorimeter/muon system after leaving no tracks in the 
inner part of the detector. If only one of the $V$'s decay after leaving the detector, the event will still appear as two $b$-jets plus MET but with additional activity due to the other $V$ now 
decaying inside the detector. 
Obviously, if both $V$'s decay inside the detector there will be no MET signature but the high $p_T$ $b$-jets can still provide the necessary trigger.  This makes the predictions of the 
current setup rather unique (due to the very large boosts as we will find below).
If both/either $V$ decays into SM particles  (\eg, $e^+e^-$, $\mu^+\mu^-$, $\pi^+\pi^-$) {\it inside} the detector it will appear as a long-lived/displaced lepton-jet\cite{ljet} and this has 
frequently been discussed as a dark sector signature (as it is here) although generally in kinematic regimes with significantly smaller boost values and without the b-tagged trigger jets. 
Searches for such lepton-jet final states 
within the context of specific signals models have been performed by both ATLAS and CMS at the LHC\cite{ljet2,new} with null results{\footnote {Note that the corresponding appearance of  
{\it prompt} lepton-jets are not very likely for the $V,S$ final states in the current setup due to the rather large boosts that are generally expected here. However, such searches {\it may}  
still play important roles in certain corners of the parameter space.}}. An important observation is that for this decay mode, the opening angle between the $V$ decay products will 
be extremely small due to the large boost that the $V$ experiences from the parent $B$ decay. Estimates of these opening angles are shown in the lower panel of Fig.~\ref{vlength} for 
the both $e^+e^-$ and $\mu^+\mu^-$ final states which represented the parameter space weighted 
averages for these expectations (and not those associated with a specific parameter choice). For the $\mu^+\mu^-$ final state case we added the additional constrain that $m_V\geq 250$ 
MeV as part of the scan while $m_e=0$ has been assumed throughout.  

\begin{figure}[htbp]
\centerline{\includegraphics[width=5.0in,angle=0]{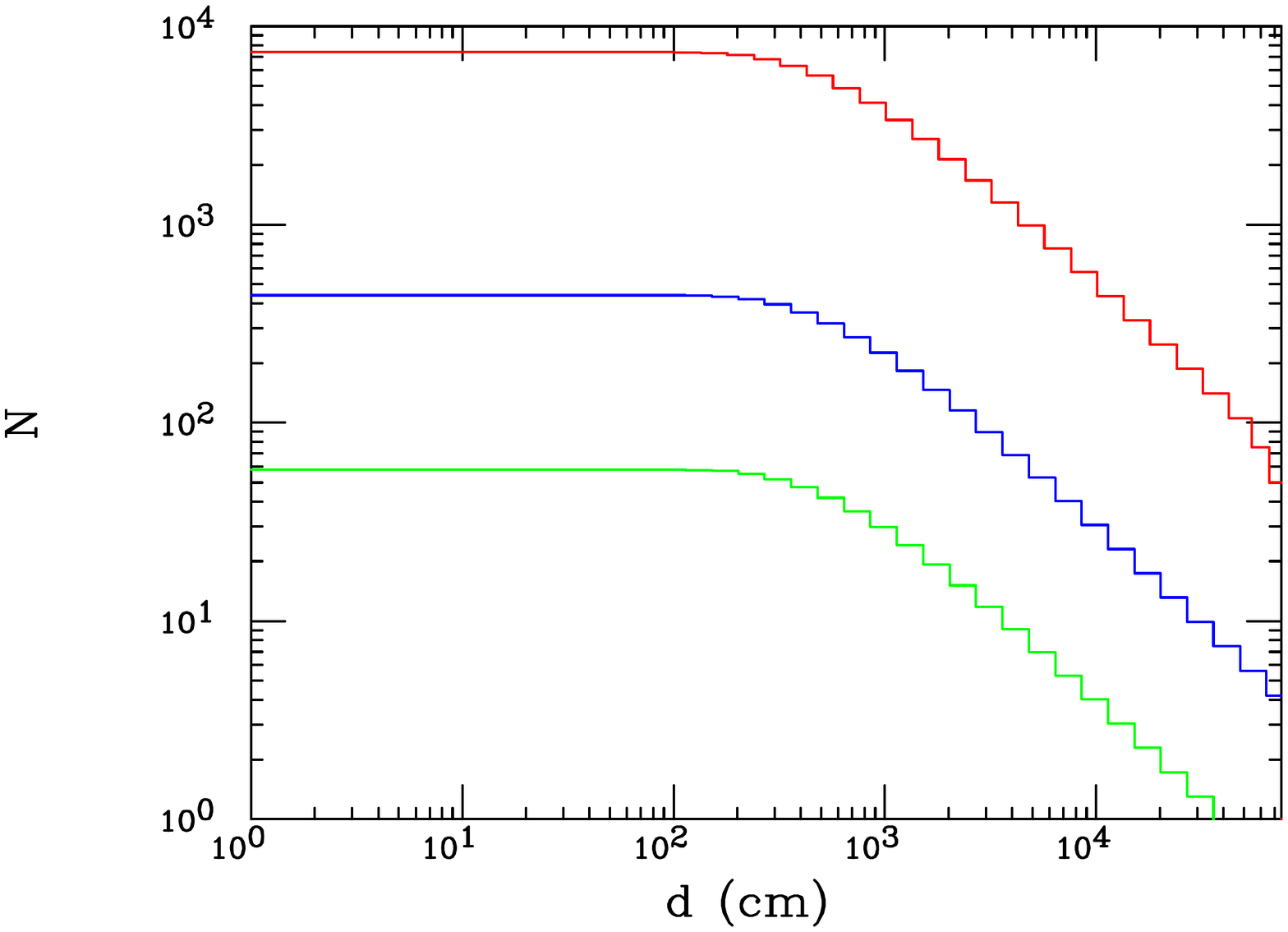}}
\vspace*{-2.5cm}
\centerline{\includegraphics[width=5.0in,angle=0]{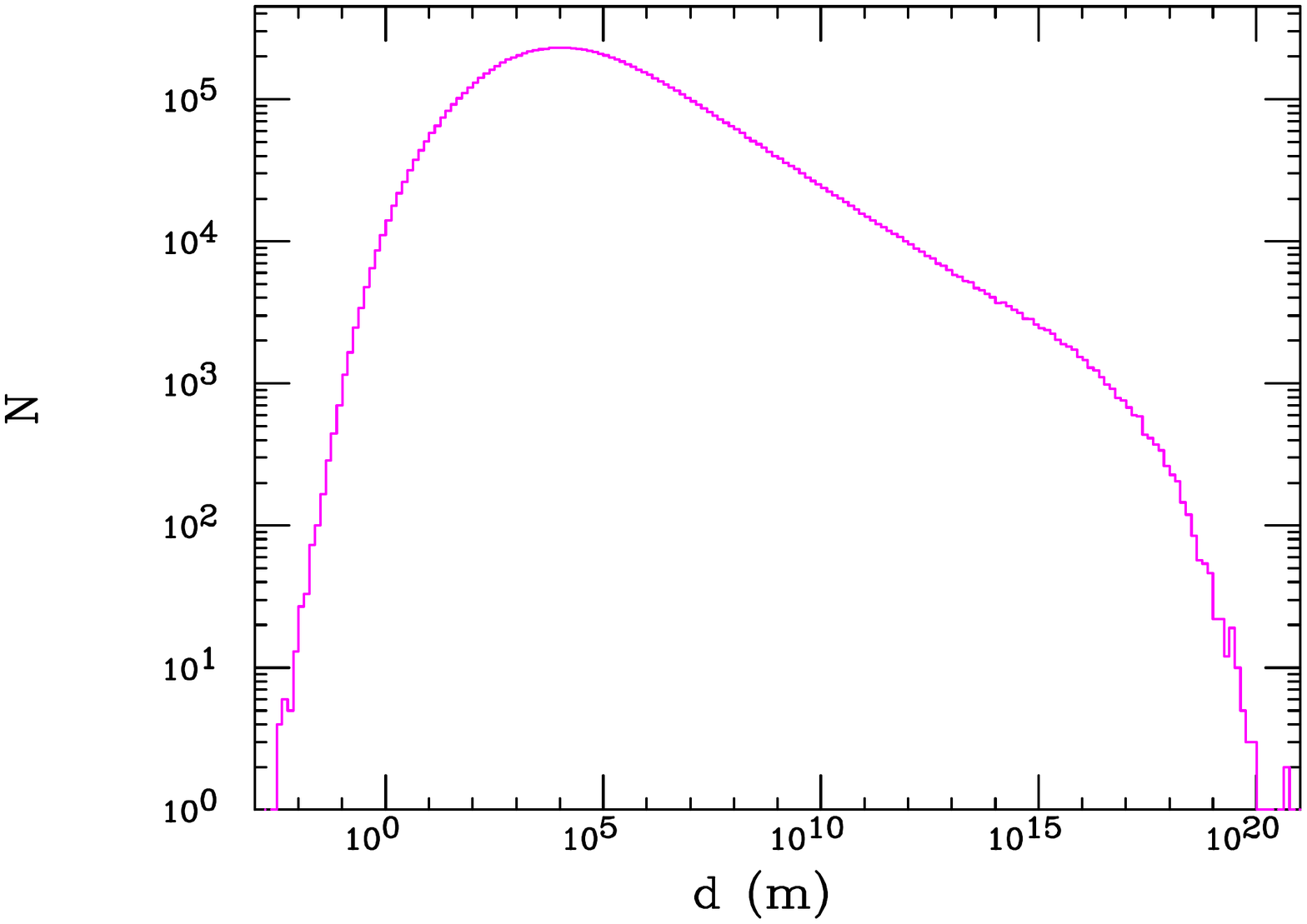}}
\vspace*{-1.50cm}
\caption{(Top) Event rate estimate for the number of single lepton-jets for the idealized ATLAS detector as discussed in the text assuming an integrated luminosity of 100 $fb^{-1}$. 
From top to bottom the histograms assume that  $m_1=1,1.5, 2$ TeV, respectively. 
(Bottom) Boosted decay length ($d$) distribution of the $S$ from the decay $B\to bS$ assuming that $m_B=1$ TeV as described in the text assuming $m_V<m_S<2m_V$. 
Note that values of $d > 10^{12}$ m or so are likely excluded by consideration of cosmological and nucleosynthesis constraints.}
\label{slength}
\end{figure}

Since displaced lepton-jets decaying at different depths in the detector 
are somewhat challenging to simulate with DELPHES\cite{ljet,DELPHES}, we can obtain an (albeit extremely crude) estimate for the {\it largest} possible rate for such events by 
assuming that, \eg, ATLAS is a perfect detector of radius $\sim$7m with 100$\%$ acceptance and efficiency so that any 
$V \to$ lepton-jet events produced inside this detector volume are observed. Assuming an integrated luminosity of 100 $fb^{-1}$ for purposes of demonstration, and knowing the $B\bar B$ 
cross section as a function of the $B$ mass from above we know the total number of $B$'s produced, all of which decay to $V$ or $S$ with an equal branching fraction.  For any given 
value of the boosted decay length, $d$, (employing the range as given in Fig.~\ref{vlength} and determined from our parameter space scan described above) 
we can determine the probability that $V$ will decay 
before reaching the $7m$ edge of the detector, correcting for the different values of $m_1$. From this information, we then know the maximum number of single 
lepton-jet events that might be observed under these extremely optimistic assumptions of perfect efficiency (here assuming both the $bV$ and $bS$ final states produce at least one 
lepton-jet) as a function of $d$. The result of this very simple calculation, which 
is clearly a sizable overestimate of reality, is shown in the upper panel of Fig.~\ref{slength}. From this result we can argue that for values of $d$ below $\sim$ 10's of meters lepton-jets 
from the decay of $B$-like portal matter in the mass range of interest may likely be observable with the Run II data set. Of course a much more realistic analysis of this possibility is 
clearly extremely necessary. 

When the decay $B\to bS$ occurs the critical issue is whether or not the bound $1<m_S/m_V<2$ is satisfied; although this appears to be a small region of the parameter space, it is the 
one most analogous to that realized in the SM for the Higgs and $W^\pm$ so we need to take this mass range possibility quite seriously. 
If the mass ratio does lie in this range, when $S$ receives a boost of $\sim 10^3-10^4$ from the 
original $B$, the boosted decay length of $S$ can easily exceed $\sim 10-100$ km as its un-boosted decay length is typically $\sim 2F^{-1}$ m as discussed above. This result is 
shown in the lower panel of Fig.~\ref{slength} which makes use of the previous parameter space scan and, in addition, now includes a scan over values of $1<m_S/m_V<2$ 
with a flat prior.  Here the generally small value of 
the function $F(r)$ plays a rather significant role in increasing both the $S$ lifetime and the corresponding boosted decay length. In such a situation it is quite likely that the 
$V$'s resulting from the $S$ decay will themselves decay to SM particles very far outside the detector.  In such cases we return to the two $b$-jet plus MET signature as previously 
discussed and 
which yields a bound of $m_{1,2}>1-1.5$ TeV or more.  We note that if $H-S$ mixing {\it is} important at the level discussed above, and if $S\to \mu^+\mu^-$ is kinematically open, then 
its un-boosted decay length will be greater than $\sim 0.55$ m; such values would still lead to quite long boosted decay lengths and result in the likely decay of $S$ outside of the detector 
much of the time. 

\begin{figure}[htbp]
\centerline{\includegraphics[width=5.0in,angle=0]{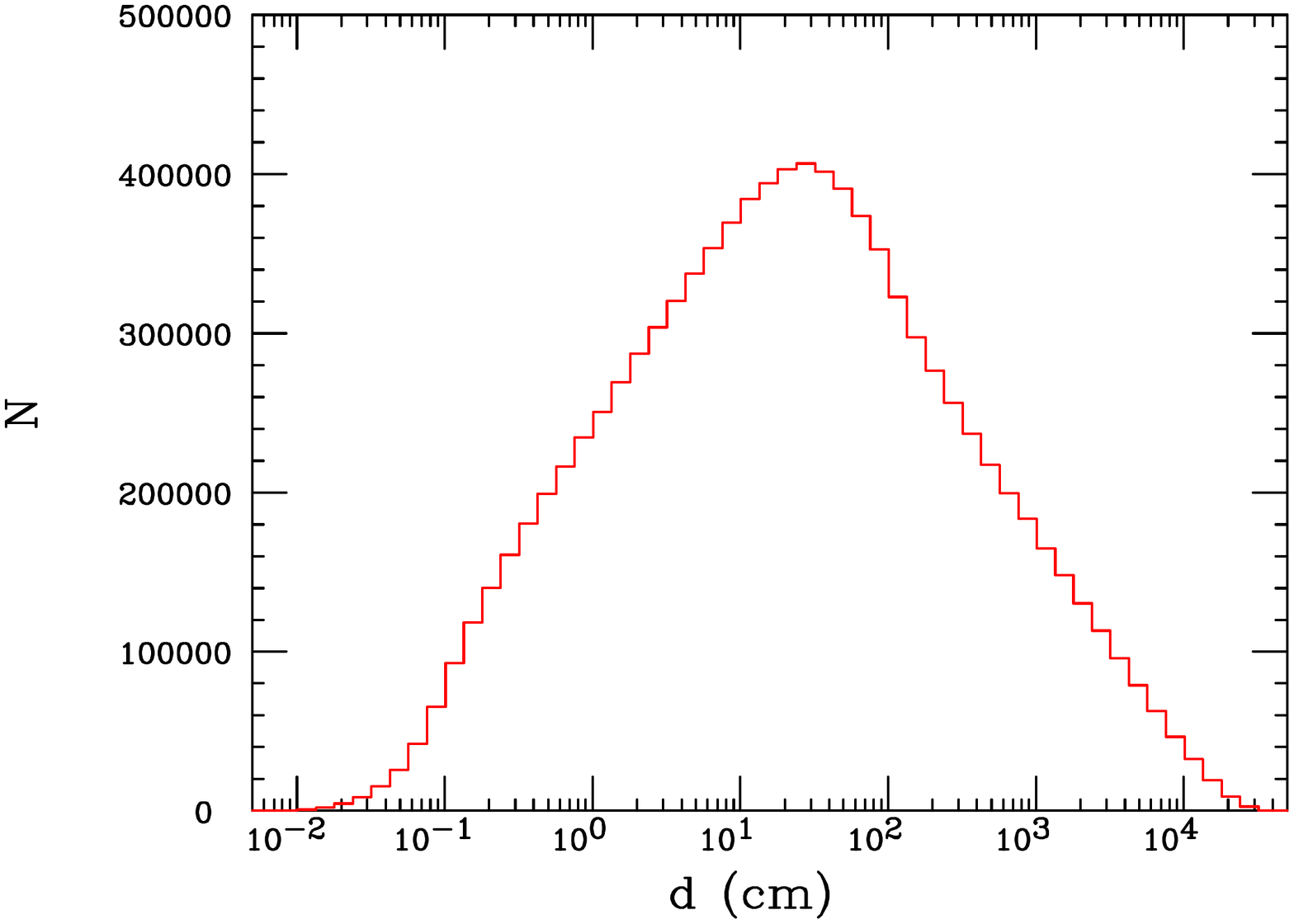}}
\vspace*{-2.5cm}
\centerline{\includegraphics[width=5.0in,angle=0]{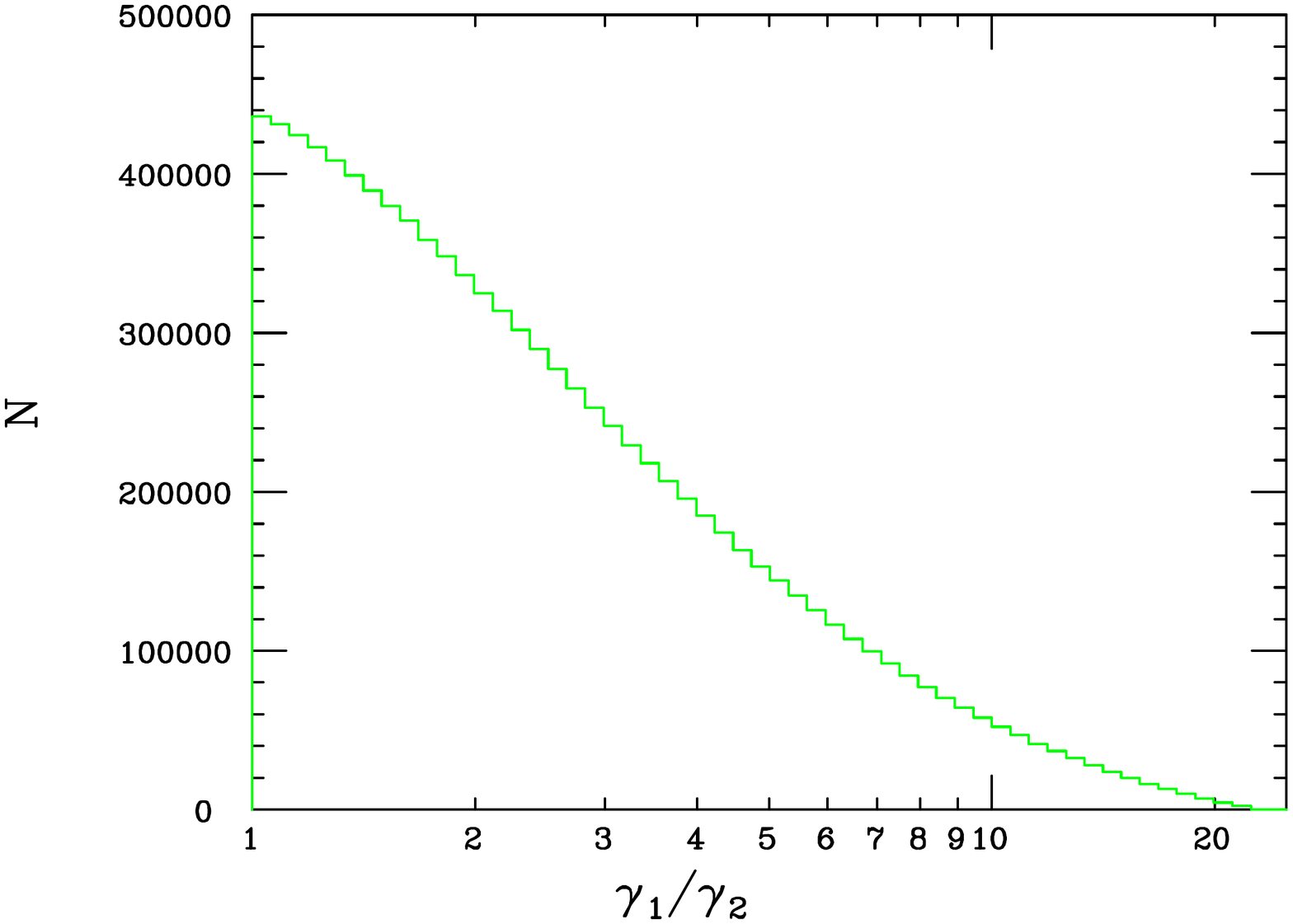}}
\vspace*{-1.50cm}
\caption{(Top) Boosted decay length ($d$) distribution of the $V$ arising from the decay chain $B\to bS, S\to VV$ assuming that $m_B=1$ TeV as described in the text.
(Bottom) Ratio of the boosts of the two $V$'s produced from a single initial $B$ (or $\bar B$) and the $B\to bS, S\to VV$ decay chain. $V_1$ is always defined to be the 
state with the larger boost.}
\label{biglife}
\end{figure}

\begin{figure}[htbp]
\centerline{\includegraphics[width=5.0in,angle=0]{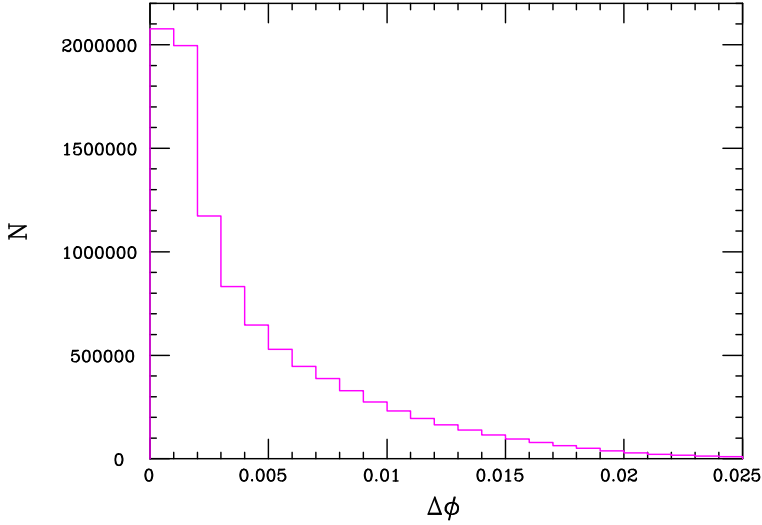}}
\vspace*{-1.50cm}
\caption{
Distribution of the opening angle between the two $V$'s in $S$ decay, $\Delta \phi$, as described in the text. 
}
\label{biglife2}
\end{figure}

When $m_S>2m_V$, on the other 
hand, the $S\to VV$ decay is prompt resulting in a two $b$-jet plus $4V$ final state with the two pairs of $V$'s from each $S$ being strongly columnated due to the large boost.  Since 
the $V$'s in such a case result from a secondary decay in the chain and their pre-boost velocities can be oriented arbitrarily (with a flat distribution with respect to the motion of the $S$ 
since it is spin-0), their actual boosts will be somewhat reduced. In this case the secondary 
$V$ boosted decay lengths, shown in the top panel of Fig.~\ref{biglife}, will be somewhat shorter than when 
the $V$ arises directly from $B\to bV$. (To obtain these numerical results we have assumed that $2\leq m_V/m_S \leq 5$ with a flat prior.) Note that some of the $V$ decays in this parameter 
space region may now appear as {\it prompt} unlike when $B\to V$ directly. This $d$ distribution is found to display a peak value of $\sim 30$ cm but, again, with rather long 
tails, extending 
below $\sim 1$ mm at the lower end. The two $V$'s from a given $S$ will likely not decay at the same time or possibly even in the same region within the detector since the ratio of the 
{\it distinct} boosts that the two $V$'s receive from the $B\to S\to VV$ decay chain can be O(1), this again due to the random orientation of the produced $V$ relative to direction of the 
moving $S$. The lower 
panel of Fig.~\ref{biglife} shows the ratio of these two boosts where $V_1$ is always defined as the one obtaining the larger boost in this decay chain. We see that while a ratio of unity 
is the most likely value, there is a large region of this model space where this ratio is significantly greater than unity. On average, employing the same scan results as above, we find that 
the opening angle between the two $V$'s in the detector is generally still quite small as seen in Fig.~\ref{biglife2}. It is to be remembered that this result represents the weighted distribution 
for this opening angle averaged over the model parameter space and is {\it not} that corresponding to a specifically chosen parameter space point.

\begin{figure}[htbp]
\centerline{\includegraphics[width=5.0in,angle=0]{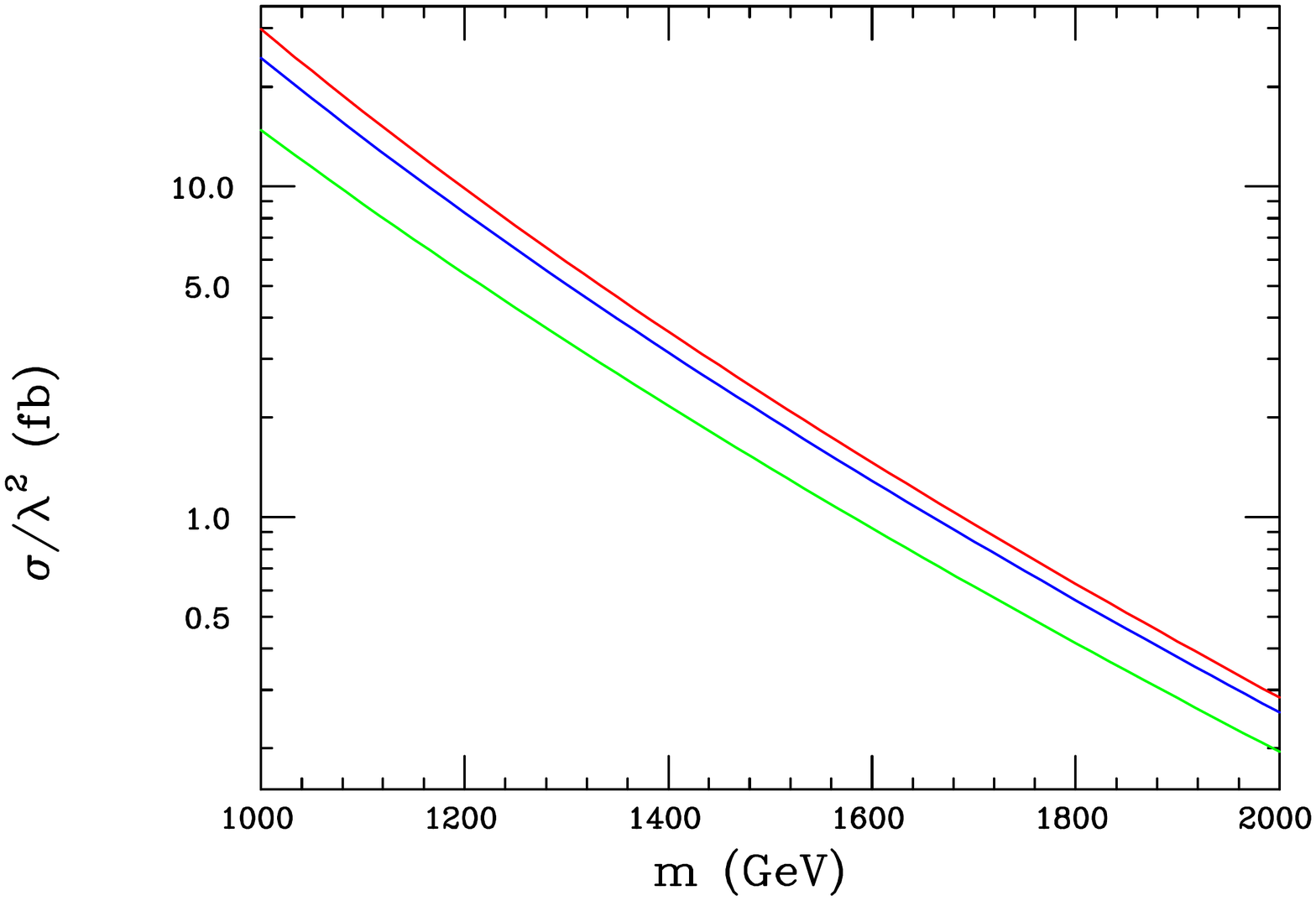}}
\vspace*{-2.5cm}
\centerline{\includegraphics[width=5.0in,angle=0]{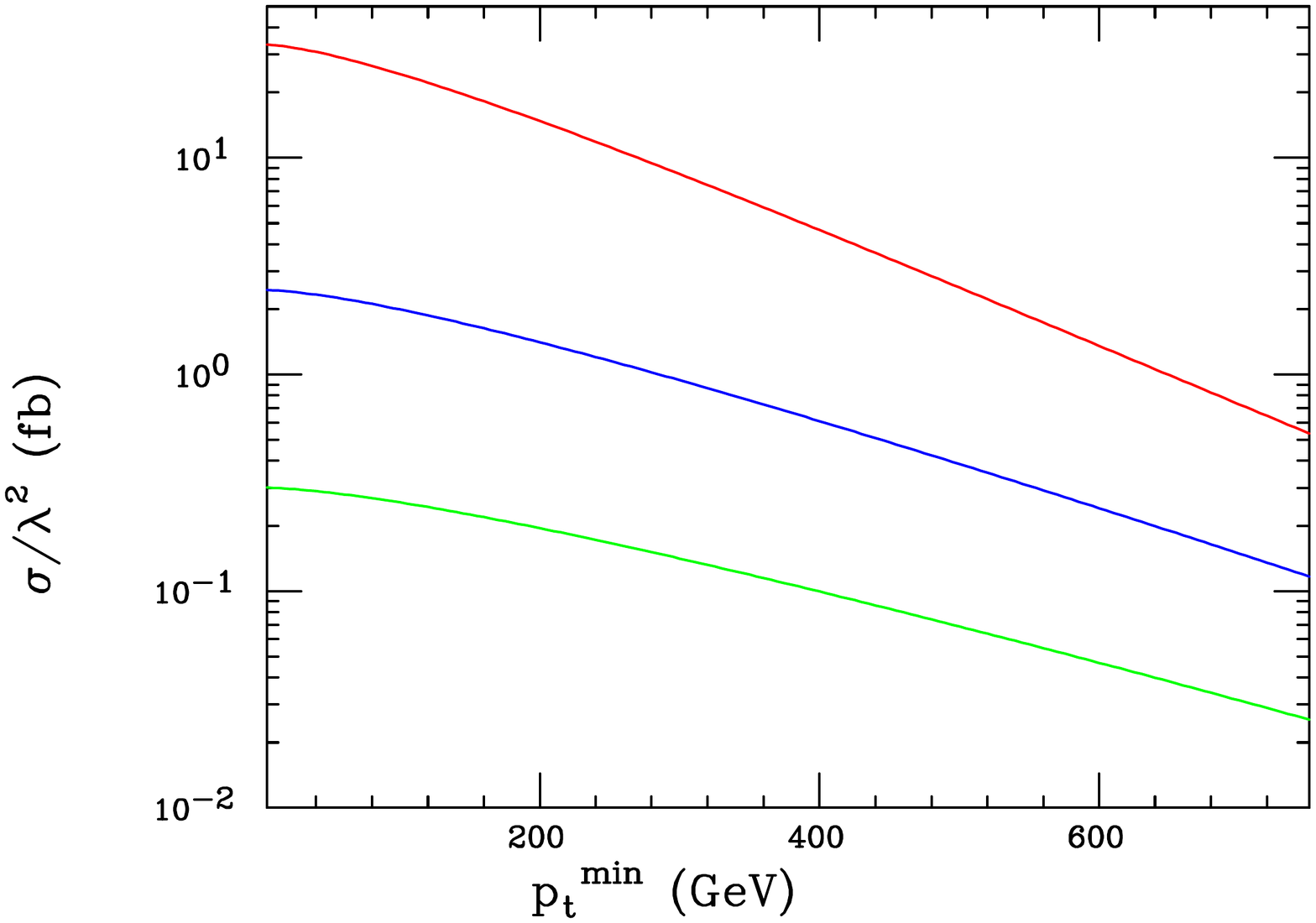}}
\vspace*{-1.50cm}
\caption{(Top) LO cross section for single $B$ production in association with $S/V$ as a function of the $B$ mass. From top to bottom a cut of $p_T(S,V)>50,100, 200$ GeV, 
respectively, has been applied. (Bottom) Same as in the top panel but now as a function of $p^{min}_T(S,V)$ for, from top to bottom, a $B$ mass of 1, 1.5, or 2 TeV, respectively, 
has been assumed. In both panels a cut of $|\eta_{V,S}|<2.5$ has also been applied.}
\label{singleb}
\end{figure}

\begin{figure}[htbp]
\centerline{\includegraphics[width=5.0in,angle=0]{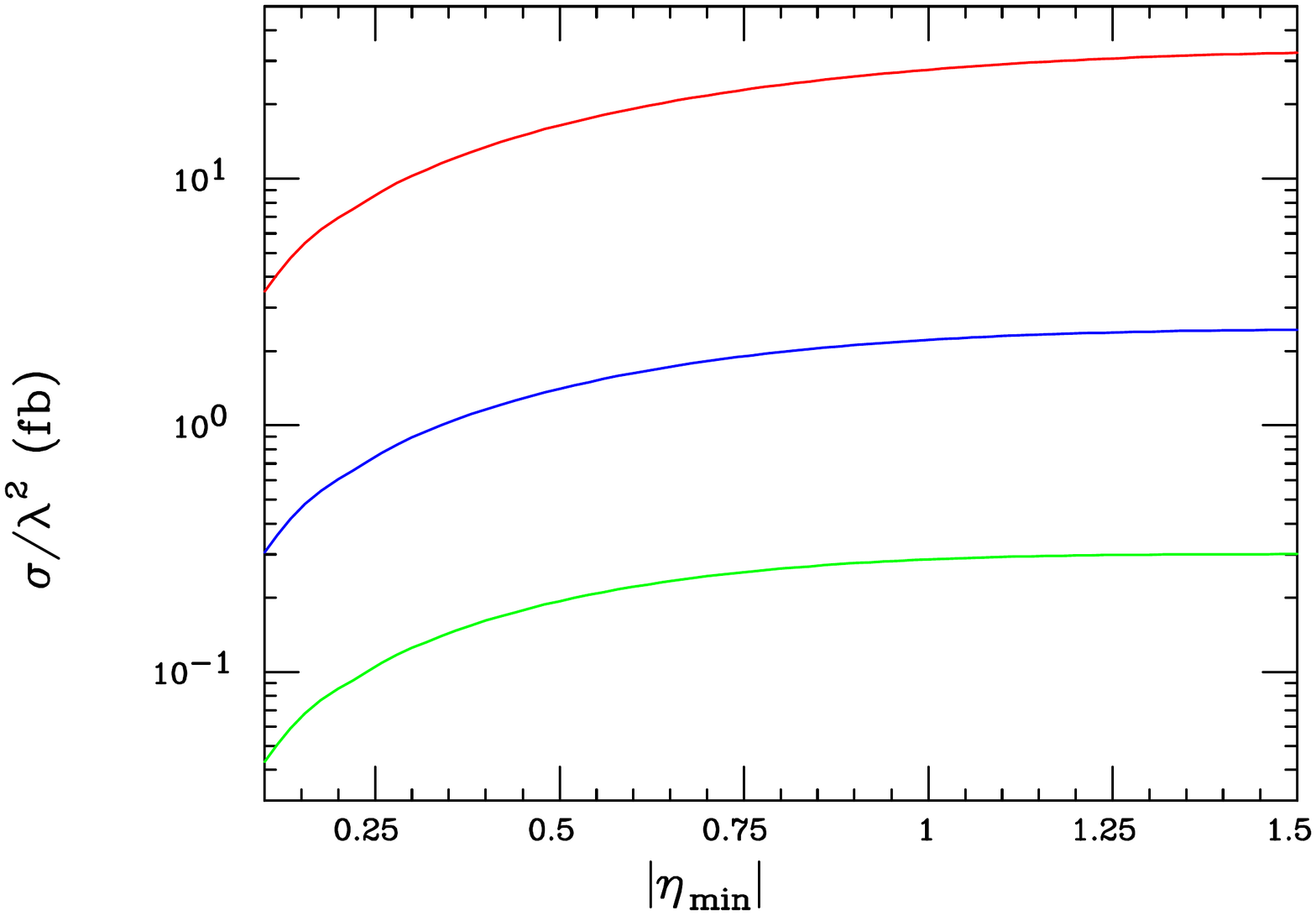}}
\vspace*{-1.50cm}
\caption{
Associated production cross section for $B+V/S$ as a function of the $|\eta_{V,S}|_{min}$; from top to bottom, we assume a $B$ mass of 1, 1.5, or 2 TeV, respectively.
}
\label{singleby}
\end{figure}

In addition to $B\bar B$ pair production, we briefly note that single $B$ associated production with either $V/S$ can also occur at a significant rate for $\lambda_i$ being O(1) with a LO 
cross section that is given by (in the limit $m^2_{b,V,S}\to 0$) 
\begin{equation}
\frac{d\sigma}{d\hat t}=\frac{\lambda^2_i \alpha_s}{48\hat s^2}\Big[\frac{(\hat s+\hat t)^2}{-\hat s t'}+m^2_i(\hat s+\hat t)(\frac{1}{\hat s^2}+\frac{1}{t'^2})-\frac{2m^2_i \hat t}{t'}(\frac{2}{t'}+
\frac{1}{s})\Big]\,,
\end{equation}
where $t'=\hat t-m^2_i$; note that in this limit, the Equivalence Theorem\cite{GBET} tells us that production with either $S$ or $V$ yields the same cross section.  As can be seen from 
this expression, the single production process occurs via $gb$ fusion ($\to B_iV/S$) with virtual $b(B_i)$ exchanges in the $s(t)-$channels. We note that in the 
corresponding case where $F_i=T_i$ such a production mechanism is absent since to a good approximation there are no tops in the proton. The value of this associated production 
cross section at the $\sqrt s=13$ TeV LHC is shown in the top panel 
of Fig.~\ref{singleb} as a function of the $B_i$ mass for different $p_T$ cuts on $S,V$ and as a function of the cut $p^{min}_T(S,V)$ for different choices of the $B_i$ mass; in all cases 
an additional cut requiring a central event,  
$|\eta_{V,S}|<2.5$, has been applied. This associated production process, as in the case of $B\bar B$ production, leads to a very highly boosted $V,S$ arising from the $B$ decay itself. 
However, the {\it associated} $V,S$, which is also produced in this process, generally arrives with lower $p_T$ and at larger rapidities, as can be seen in the lower panel of 
Fig.~\ref{singleb} and in Fig.~\ref{singleby}. Thus for fixed $p_T$ this $V,S$ is likely to experience a somewhat smaller boost especially at the smaller rapidities, $|\eta_{V,S}|\to 0$, 
in comparison to those produced in the $B$ decay. In particular we see that, for the range of masses of interest to us, almost all the events are found to lie in the region 
$|\eta_{V,S}| \gsim 0.5-1$. Such events can be triggered on by the appearance of the single high $p_T$ b-jet which now appears in the final state. In the case where $V/S$ decay outside 
the detector the event will appear as MET plus a single $b$-jet for which there have so far been only null searches at the LHC~\cite{oneb} with a current bound of $\sim1$ fb on the 
production cross section (after suitable cuts).

Finally we remind the reader 
that the possibility of the $\bar BB\to \bar bb+SS/VV/VS$ final state at the LHC is just one example of what may happen for any of the $\bar FF$ portal matter final states discussed 
above. As mentioned earlier one should also insure that the $\bar tt+S,V$ final state is also explored since it is qualitatively similar to the case of $F=B$ except that top reconstruction 
is somewhat less efficient than is the tagging of high $p_T$ $b$-jets. In the case of $F=T$, the top quark polarization in the $T\to tV,S$ decays can be used to obtain an additional 
handle on the chirality structure of the $TtV$ and $TtS$ vertices\cite{Zhou:2019alr} which can yield important information about the portal matter multiplet structure, in particular, allowing one to 
distinguish isodoublet from isosinglet scenarios.  Also, one should not ignore the pair production of the {\it leptonic} MP states at the LHC through their electroweak 
interactions which can only have sizable rates at somewhat lower masses and that can lead to MET +$V,S$ when $F=N$ or opposite sign, same flavor leptons $+V,S$ final states when 
$F=E,M$.

\section{Summary and Conclusions}

The generation of the kinetic mixing between the dark photon and the hypercharge gauge boson of the Standard Model is a necessary element in the construction of the dark photon 
mediator model.  To accomplish this KM requires the existence of portal matter simultaneously carrying both dark and SM charges and if this matter is fermionic, as we argue above, it 
must be vector-like with respect to the SM 
to avoid anomalies and the experimental constraints from arising from both precision electroweak data and Higgs decays. Among the set of possible SM 
transformation properties for these states, one can only chose from among those that allow for the portal matter to decay. This being the case, we demonstrated that the only way to do this at 
tree-level, given that the dark Higgs is a SM singlet, is to require that the portal matter transform as vector-like versions of the SM fields themselves. Of these five possible states, those that are 
also singlets under the $SU(2)_W$ gauge group presents us with the simplest possibilities and it is these cases that we have mainly focused on here. However, it is quite likely that in 
more UV-complete frameworks the portal matter sector may be some more complex and may be a more interesting combination of these different vector-like states. In particular, we have found
 that if the portal matter and the dark Higgs, which is responsible for generating the mass of the dark photon, carry the same dark charge, then the vev of the dark Higgs will induce a mixing between 
 the portal matter and the analogous SM field carrying the same SM quantum numbers thus allowing for the desired (prompt) decay path. This same mixing leads to parity violating dark photon 
 couplings to (only) the specific SM fields which are the conventional analogs of the portal matter. If the portal matter transforms like the $e_R$, these parity-violating interactions may be probed by, 
 \eg, the MOLLER experiment. Similarly, if the portal matter transforms as $\mu_R$, this same mixing will also lead to potential new contributions to the muon's g-2. This mixing is, in particular, 
also responsible for the specific prompt portal matter decays $F\to fV,S$, which are the (by far) dominant modes for these new states, that can be observable at the LHC and future colliders. As 
we have shown, although these portal matter fields are vector-like fermions similar to the ones most commonly discussed in the literature, they do not decay into the familiar and usually sought $H,W^\pm$ or $Z$ 
final states with any significant branching fractions so that these final states are {\it not} the ones to be employed in searching for portal matter. Here we discussed at some length in the analysis 
above the specific case when the portal matter transforms as a color triplet, \ie, similar to $b_R$ in the SM, which leads, in the case of portal matter pair production, to b-jet pairs (which can be used to 
trigger on these events) plus possible MET in a manner similar to (but with larger cross sections than) bottom squarks, or, more interestingly, b-jets plus very highly boosted displaced 
lepton-jet signals at the LHC. In most of the model parameter space these 
lepton-jets are much more highly boosted than what usually results from the signal models that have been previously employed in all the LHC searches and  can pose a serious 
resolution challenge for collider detectors. Qualitatively similar signatures are possible if the portal matter transforms like $t_R$ except for the presence of boosted top jets instead of $b-$jets 
in the final state.  Also in the case of a $b_R$-like portal matter (but not in the $t_R$-like case), the O(1) couplings to the dark Higgs allow for the possibility that single production of portal matter may occur 
at significant rates with a unique signature having only a single triggering jet. 

The exploration of the physics of portal matter as discussed above, especially at colliders, is just at its beginning stages and there are multiple directions which can be pursued which 
we leave for future study.

\section*{Acknowledgments}
The author would like to particularly thank J.L. Hewett and T.D. Rueter for discussions. He would also like to especially the members of the SLAC ATLAS Collaboration,  
in particular, Su Dong, C. Young, M. Diamond, R. Bartoldus, M. Kagan, C. Vernieri and A. Schwartzman,  
for very valuable discussions on the long-lived particles and lepton-jets analyses with ATLAS. This work was supported by the Department of Energy, Contract DE-AC02-76SF00515.

\section*{Appendix: Coupling and Interaction Summary}

We collect and review in this Appendix some useful approximate expressions for the couplings of the various physical fields that we have employed in the present analysis. 

Recall that we assume that the dark photon has a mass in (roughly) the range $\sim 0.1-1$ GeV implying that the singlet dark Higgs vev is $v_s\lsim $ a few GeV; the portal matter-SM 
dark Higgs couplings, $\lambda_i$, are assumed to be $O(1)$. Correspondingly, a pair of mass parameters, $X_i=\lambda_i v_s/\sqrt 2 \sim$ GeV can also be defined for convenience. 
Give the experimental constraints and future experimental interest, we have consequently focussed on KM parameter values of $\epsilon \sim 10^{-(3-5)}$. Since portal matter fields 
are generally much heavier than their corresponding SM partners, of order several hundred GeV to more than 1 TeV, to leading order and for the isosinglet models we consider, we can treat 
the left-handed fermion mixing matrix as approximately diagonal, \ie, $U_L \simeq I$ and deal only with a non-trivial $U_R$ in our phenomenological analyses. Since $U_L \simeq I$, 
one finds that the decays of portal matter fields into their SM partners and the SM fields $W^\pm,Z,H$ are relatively suppressed.  As is usual, all the SM fields have an interaction with 
the dark photon given by $e\epsilon Q_{em}$.  However, the non-trivial $U_R$ mixing induced by the dark Higgs does three things: ($i$) it leads to an additional contribution, corresponding 
to a parity violating, right-handed coupling, of the dark photon to the SM field, $g_D(C_R)_{11}\simeq g_D(X_1/m_1)^2$, but {\it only} for the SM field which has this specific portal matter 
partner (here of mass $m_1$). A non-trivial $(C_R)_{11}$ leads to all of the leptonic portal matter discussion above involving the dark photon. We note that $g_D(C_R)_{11}$ is 
expected to be not very different than $e\epsilon$ in magnitude and so it is useful to define a relative strength parameter $y=g_d(C_R)_{11}/2e\epsilon$.  ($ii$) This same non-trivial 
mixing matrix also leads to an off-diagonal coupling of the dark photon between this portal field and its SM partner, \eg, $g_D(C_R)_{12} \simeq g_DX_1/m_1$, allowing for a new 
contribution to g-2, for the portal matter to decay and providing, \eg,  a production signature at the LHC as discussed above. Finally, ($iii$) since this portal matter-SM mixing is induced 
by an off-diagonal coupling of the dark Higgs, this off-diagonal interaction remains in the mass eigenstate basis, \eg, $\beta_{21}\simeq \lambda _1 X_1/(\sqrt 2 m_1)$, which allows 
for the dark Higgs couplings contributing to $g-2$ and for a second decay channel of the portal matter fields at the LHC.

\vspace*{1.50cm}


\end{document}